\documentclass[journal,final]{IEEEtran}
\usepackage{url}

\usepackage{style/jurestyle}
\usepackage{style/juremath}
\usepackage{style/jurefigurestyle}

\newcommand{\review}[1]{{\color{black}#1}}

\usepackage{standalone}
\usepackage{tikz}

\newtheorem{theorem}{Theorem}
\newtheorem{corollary}{Corollary}
\newtheorem{lemma}{Lemma}

\newtheorem{example}{Example}

\begin{document}

\title{Mismatch in the Classification of Linear Subspaces: Sufficient Conditions for Reliable Classification}

%
%
%
\author{Jure~Sokoli\'c,~\IEEEmembership{Student Member,~IEEE,} Francesco~Renna,~\IEEEmembership{Member,~IEEE,} Robert~Calderbank,~\IEEEmembership{Fellow,~IEEE,} and~Miguel~R.~D.~Rodrigues,~\IEEEmembership{Senior Member,~IEEE} 
\thanks{This paper was presented in part at the 2015 IEEE International Symposium on Information Theory.}
\thanks{The work of Jure Sokoli\'c, Francesco Renna and Miguel R. D. Rodrigues was supported in part by EPSRC under grant EP/K033166/1. The work of Robert Calderbank was supported in part by AFOSR under grant FA 9550-13-1-0076 and by NGA under grant HM017713-1-0006.}
\thanks{J.~Sokoli\'c, F.~ Renna and M. R. D. Rodrigues are with the Department of Electronic and Electrical Engineering, University College London, United Kingdom 
({\sf email: \{jure.sokolic.13, f.renna, m.rodrigues\}@ucl.ac.uk}).}
\thanks{R. Calderbank is with the Department of Electrical and Computer Engineering, Duke University, NC, USA
({\sf email: robert.calderbank@duke.edu}).}
}

\maketitle


\begin{abstract}  
This paper considers the classification of linear subspaces with mismatched classifiers. In particular, we assume a model where one observes signals in the presence of isotropic Gaussian noise and the distribution of the signals conditioned on a given class is Gaussian with a zero mean and a low-rank covariance matrix. We also assume that the classifier knows only a mismatched version of the parameters of input distribution \emph{in lieu} of the true parameters. By constructing an asymptotic low-noise expansion of an upper bound to the error probability of such a mismatched classifier, we provide sufficient conditions for reliable classification in the low-noise regime that are able to sharply predict the absence of a classification error floor. Such conditions are a function of the geometry of the true signal distribution, the geometry of the mismatched signal distributions as well as the interplay between such geometries, namely, the principal angles and the overlap between the true and the mismatched signal subspaces. Numerical results demonstrate that our conditions for reliable classification can sharply predict the behavior of a mismatched classifier both with synthetic data and in a motion segmentation and a hand-written digit classification applications.
\end{abstract}

\begin{IEEEkeywords}
	Classification, mismatch, linear subspace, Maximum-a-Posteriori classifier, error floor.
\end{IEEEkeywords}

%
\IEEEpeerreviewmaketitle



\section{Introduction} 
Signal classification is a fundamental task in various fields, including statistics, machine learning and computer vision. One often approaches this problem by leveraging the Bayesian inference paradigm, where one infers the signal class from signal samples or measurements based on a model of the joint distribution of the signal and signal classes \cite[Chapter 2]{Duda2012}. 

Such joint distribution is typically inferred by relying on pre-labeled data sets. However, in practical applications, the methods used to estimate the distributions from training data inevitably lead to signal models that are not perfectly matched to the underlying one. This can be due to an insufficient number of labeled data, the  noise in the pre-labeled data  \cite{Frenay2014}, \cite{Natarajan2013, Bootkrajang2012}, or due to the non-stationary  statistical behaviour~\cite{Quionero-Candela2009}. 
 
It is therefore relevant to ask the question:

\vspace{0.1cm}
\emph{What is the impact that a mismatched classifier, i.e. a classifier that infers the signal classes based on an inaccurate model of the data distribution in lieu of the true underlying data distribution, has on classification performance?}
\vspace{0.1cm}

We answer this question for the scenario where the data classes are constrained to lie approximately on a low-dimensional linear subspace embedded in the high-dimensional ambient space. Indeed, there are various problems in signal processing, image processing and computer vision that conform to such a model, some of which are:
\begin{itemize}
\item \emph{Face Recognition}: It can be shown that, provided that the Lambertian reflectance assumption is verified, the set of images taken from the same subject under different lighting conditions can be well approximated by a low-dimensional linear subspace embedded in the high-dimensional space \cite{Basri2003}. This is leveraged in several face recognition applications \cite{Lee2005, Wright2009 , Zhang2010}. 

\item \emph{Motion Segmentation}: It can also be shown -- under the assumption of the affine projection camera model -- that the coordinates of feature points associated with rigidly moving objects through different video frames lie in a 4 dimensional linear space  \cite{vidalTutorial}, \cite{Tomasi1992}, \cite{Boult1991}. This is leveraged in  \cite{vidalTutorial} to design subspace clustering algorithms that can perform motion segmentation.

\item In general,  (affine) subspaces or unions of (affine) subspaces can also be used to model other data such as images of handwritten digits \cite{chen2010a}.

\end{itemize}

Our contributions include:
\begin{itemize}
	\item We derive an upper bound to the error probability associated with the mismatched classifier for the case where the distribution of the signal in a given class is Gaussian with zero-mean and low-rank covariance matrix.
	\item We then derive sufficient conditions for reliable classification in the asymptotic low-noise regime. Such conditions are expressed in terms of the geometry of the true signal model, the geometry of the mismatched signal model and the interaction of these geometries (via the principal angles associated with the subspaces of the true and mismatched signal models as well as the dimension of the intersection of such subspaces).
	\item We finally provide a number of results, both with synthetic and real data, that show that our sufficient conditions for reliable classification are sharp. In particular, we also use our theoretical framework to determine the number of training samples needed to achieve reliable classification in a motion segmentation and a hand-written digit classification applications. 
\end{itemize}

\subsection{Related Work}

The concept of model mismatch has been widely explored by the information theory and communication theory communities. For example, in lossless source coding problems, mismatch between the distribution used to encode the source and the true distribution is shown to lead to a compression rate penalty which is determined by the Kullback-Leibler (KL) distance between the mismatched and \mbox{the true distributions \cite[Theorem 5.4.3]{cover2006}}. 

In channel coding problems, mismatch has an impact on the reliable information transmission rate that has been characterized via inner and outer bounds to the achievable rate and error exponents of different channel models \cite{Merhav1994,Lapidoth1998, Ganti2000,Scarlett2014,Somekh-Baruch2015}. The problem of mismatched quantization is considered in \cite{Gray2003}.

The concept of mismatch has also been explored in the machine learning literature \cite{Quionero-Candela2009}. In particular,  \cite{Quionero-Candela2009} studies the impact on classification performance of training sets consisting of biased samples of the true distribution, expressing classification error bounds as a function of the sample bias severity and type.  
The effect of label noise in the training sets is also considered in classification algorithms such as Support Vector Machines \cite{Natarajan2013} and Logistic Regression classifiers \cite{Bootkrajang2012}. See also \cite{Frenay2014} for an overview of the literature on classification in presence of label noise. 




Signal classification and estimation using mismatched models is also considered in \cite{Kazakos1982, Schluter2001, Schluter2013, Verdu2010}. For example, \cite{Schluter2013} expresses bounds to the error probability in the presence of mismatch via the $f$-Divergence between the true and mismatched source distributions, and \cite{Verdu2010} expresses the mean-squared error penalty in presence of mismatch in terms of the derivative of the KL distance between the true and the mismatched distributions with respect to the decoder signal to noise ratio (SNR). \review{In particular, the work in \cite{Schluter2013} is closely related to our work in the sense that it also establishes bounds to the error probability in the presence of mismatch. The bounds presented in \cite{Schluter2013} are more general since they do not assume a particular form of probability density functions. Our work, on the other hand, leverages the assumption that signals are contained in linear subspaces in order to derive an upper bound that sharply predicts the presence or absence of an error floor. The bounds in \cite{Schluter2013} fail to capture the presence or absence of an error floor when specialized to the proposed signal model.
}

\subsection{Organization}
The remainder of this paper is organized as follows: Section \ref{sec:problemstatement} introduces the observation and signal models, the Mismatched Maximum-a-Posteriori (MMAP) classifier and the geometrical quantities associated with the signal and the mismatched model that are essential for the description of the MMAP classifier performance. The upper bound to the error probability associated with the MMAP classifier and the asymptotic expansion, which provide sufficient conditions for reliable classification in the low-noise regime, are given in Section \ref{sec:conditions}. In Section \ref{sec:experiments} the theoretical results are validated via numerical experiments. Applications of the proposed bound in a motion segmentation task and in a hand-written digit classification task are given in Section \ref{sec:applications}. The paper is concluded in Section \ref{sec:conclusions}. The proofs of the results are given in the Appendix.

\subsection{Notation}
We use the following notation in the sequel: matrices, column vectors and scalars are denoted by boldface upper-case letters ($\bX$), boldface lower-case letters ($\bx$) and italic letters ($x$), respectively.
$\bI_{N} \in \Rnn$ denotes the identity matrix and $\bZero_{M\times N} \in \Rmn$ denotes the zero matrix. The subscripts are omitted when the dimensions are clear from the context. $\be_{k}$ denotes the $k$-th basis vector in $\Rn$. The transpose, rank and determinant operators are denoted as $(\cdot)^{T}$, $\rank(\cdot)$ and $|\cdot|$, respectively. $\| \bx \|$ denotes Euclidean norm of the vector $\bx$ and $\| \bX \|_{2}$ denotes the  spectral matrix norm of the matrix $\bX$. The image of a matrix is denoted by $\im(\cdot)$ and the kernel of a matrix is denoted by $\ke(\cdot)$. The sum of subspaces $\mathcal{A}$ and $\mathcal{B}$ is denoted as $\mathcal{A} + \mathcal{B}$ and the orthogonal complement of $\mathcal{A}$ is denoted as $\mathcal{A}^\perp$. $\log(\cdot)$ denotes the natural logarithm, and the multi-variate Gaussian distribution with the mean $\bmu$ and covariance matrix $\bSigma$ is denoted as $\N(\bmu, \bSigma)$. We also use the following asymptotic notation: $f(x) = \cO(g(x))$ if $\lim_{x \rightarrow \infty} \frac{f(x)}{g(x)} = c$, where $c>0$, and  $f(x) = o(g(x))$ if $\lim_{x \rightarrow \infty} \frac{f(x)}{g(x)} = 0$.

\section{Problem Statement} \label{sec:problemstatement}

We consider a standard observation model:
\begin{IEEEeqnarray}{c}
	\by = \bx + \bn  \label{eq:measurement}
\end{IEEEeqnarray}
where $\by \in \Rn$ represents the observation vector,  $\bx \in \Rn$ represents the signal vector and $\bn \sim \N(\bZero, \sn \bI) \in \Rn$ represents observation noise, where $\sn$ denotes the noise variance per dimension.\footnote{This noise vector can also model the fact that data does not always lie exactly on a low-dimensional subspace but rather approximately on a low-dimensional subspace \cite{chen2010a}.}
 We also assume that the signal $\bx \in \Rn$ is drawn from a class $c \in \{1, \ldots, C \}$ with prior probability $P(c=i) = p_{i}$, and that the distribution of the signal $\bx$ conditioned on a given class $c=i$ is Gaussian with mean zero and \mbox{(possibly) low-rank covariance matrix $\bSigma_{i} \in \Rnn$, i.e.}
\begin{IEEEeqnarray}{c}
	\bx | c = i \sim \N(\bZero, \bSigma_{i}) \,,   \label{eq:sigmodel}
\end{IEEEeqnarray} 
with $\rank(\bSigma_{i}) = r_{i}  \leq N$. Therefore, conditioned on a given class $c=i$, the signal lies on the linear subspace spanned by the eigenvectors associated with the positive eigenvalues of the covariance matrix $\bSigma_{i}$.

The classification problem involves inferring the correct class label $c$ associated with the signal $\bx$ from the signal observation $\by$. It is well known that the optimal classification rule, which minimizes the error probability, is given by the Maximum-A-Posteriori (MAP) classifier \cite[Chapter 2.3]{Duda2012}:
\begin{IEEEeqnarray}{c}
	\ha{c} = \argmax_{i \in \{1, \ldots, C \}} p(c = i | \by) = \argmax_{i \in \{1, \ldots, C \}} p(\by | c = i ) p_{i}  \,, \label{eq:MAP}
\end{IEEEeqnarray}
where $p(c = i| \by)$ represents the \textit{a posteriori} probability of class label $c=i$ given the observation $\by$ and 
\begin{IEEEeqnarray}{c}
	p(\by | c = i) =  \oneo{\sqrt{(2 \pi)^{N} | \bSigma_{i} + \sn\bI|}} e^{-\oneo{2} \by^{T} (\bSigma_{i} + \sn \bI)^{-1} \by} \,
\end{IEEEeqnarray}
represents the probability density function of the observation $\by$ given the class label $c = i$.

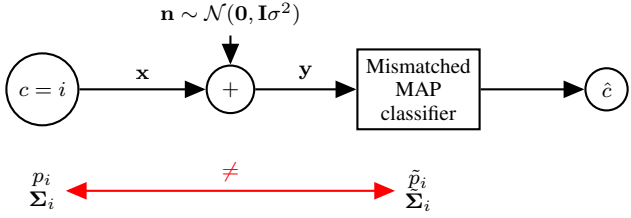
\begin{figure}[t]
	\centering
	\begin{tikzpicture}[auto, thick, node distance=2.5cm, >=triangle 45,font=\footnotesize]
	\draw
		node at (0,-5) [draw, circle] (mclass) {$c = i$}
		node [below of=mclass, rectangle, style={align=center}, node distance = 1.35cm](mparam1) {$p_{i}$ \\ $\mathbf{\Sigma}_{i}$}
		node [right of=mclass, circle, draw] (msum1) {$+$}
		node [above of= msum1, rectangle,node distance = 1cm] (mnoise) {$\mathbf{n} \sim \mathcal{N}(\mathbf{0}, \mathbf{I} \sigma^{2})$}
		node [right of=msum1, rectangle, draw, style={align=center}] (mMAP) {Mismatched \\ MAP \\classifier}
		node [right of=mMAP, circle, draw] (mMAPclass) {$\hat{c}$}
		node [below of=mMAP, rectangle, style={align=center}, node distance = 1.35cm] (mparam2) {$\tilde{p}_{i}$ \\ $\tilde{\mathbf{\Sigma}}_{i}$};
	
		\draw[->] (mclass) -- node {$\mathbf{x}$} (msum1);
		\draw[->] (msum1) -- node {$\mathbf{y}$} (mMAP);	
		\draw[->] (mMAP) -- node {} (mMAPclass);
		\draw[->] (mnoise) -- node {} (msum1);	
		\draw[,<->,red] (mparam1) -- node {$\neq$} (mparam2);		
	\end{tikzpicture}

		\vspace*{-7pt}
	\caption{System Model}
	\label{fig:system}
		\vspace*{-7pt}
\end{figure}

However, we assume that the classifier does not have access to the true signal parameters $p_{i}$, $i = 1, \ldots, C$ and  $\bSigma_{i}$, $i = 1, \ldots, C$ but rather to a set of mismatched parameters  $\ti{p}_{i}$, $i = 1, \ldots, C$ and  $\tbSigma_{i}$, $i = 1, \ldots, C$, where $\ti{p}_{i}$ is the  mismatched \emph{a priori} probability of the $i$-th class and $\tbSigma_{i}$ is the mismatched covariance matrix associated with the class $i$ with $\rank(\tbSigma_{i}) = \ti{r}_{i}  \leq N$.\footnote{We assume that $C$ and $\sn$ are known. Since we study the scenario where $\sn \to 0$, the assumption that $\sn$ is known exactly is immaterial.} (See Fig. \ref{fig:system}.)

Such a Mismatched-MAP (MMAP) classifier delivers the class estimate\begin{IEEEeqnarray}{c}
	\ti{c} = \argmax_{i \in \{1, \ldots, C \}}  \ti{p}(c=i | \by) = \argmax_{i \in \{1, \ldots, C \}}  \ti{p}(\by | c = i) \ti{p}_{i}\,, \label{eq:MMAP}
\end{IEEEeqnarray}
where $\ti{p} (c=i|\by)$ denotes the mismatched \emph{a posteriori} probability of class label $c=i$ given observation $\by$ and
\begin{IEEEeqnarray}{c}
	\ti{p}(\by | c = i) = \oneo{\sqrt{(2 \pi)^{N} | \tbSigma_{i} + \sn\bI|}} e^{-\oneo{2} \by^{T} (\tbSigma_{i} + \sn \bI)^{-1} \by}
\end{IEEEeqnarray} 
denotes the mismatched probability density function of the observation $\by$ given the class label $c = i$. 

The probability of error associated with a MMAP classifier is given by:
\begin{IEEEeqnarray}{c}
	P (e) = \sum_{i = 1}^{C} p_{i} \cdot P (e | c=i) \label{eq:MMAPerror}
\end{IEEEeqnarray}
where
\begin{IEEEeqnarray}{rCl}
	P (e | c=i) &=& \Iii  {p} (\by | c = i)  \nonumber \\
	 & & 	 \cdot u \left( \max_{\substack{j \\ j\neq i}} \log  \left( \frac{ \tilde{p}_j \tilde{p}(\by|c=j) }{ \tilde{p}_i \tilde{p}(\by|c=i)} \right) \right) \, \dd\by \label{eq:MMAPerrori}
\end{IEEEeqnarray}
and $u(\cdot)$ is the unit-step function. This error probability cannot be calculated in closed form, but it can be easily bounded.

Our goal is to study the performance of the MMAP classifier by establishing conditions, which are a function of the geometry of the true and mismatched signal models as well as the interaction of such geometries, for reliable classification in the low-noise regime i.e. such that $\lim_{\sigma^2 \to 0} P(e) = 0$.

\subsection{Geometrical Description of the Signals} \label{sec:geom}

\review{
Our characterization of the performance of the MMAP classifier will be expressed via various quantities that embody the geometry of the true signal model, the geometry of the mismatched signal model, and their interplay. The quantities central to the analysis are given in Table~\ref{tab:quantities} and the relationships between the presented quantities are summarized in Table~\ref{tab:relationships}.

\begin{table*}[t]
\caption{Main quantities used in the analysis} \label{tab:quantities}
\centering
\begin{tabular}{lcl}
Subspace & Dimension & Description  \\
\hline 
$\im( \bU_i )$ & $ r_i$ & signal space of class $i$ \\
$\im(\tbU_i )$ & $\ti{r}_i$ & mismatched signal space of class $i$ \\
$\im(\tbU_{ij}^\cap )$ & $\ti{r}^\cap_{ij}$ & intersection of the mismatched signal spaces of classes $i$ and $j$  \\
$\im(\tbU_{ij}' )$ & $ \ti{r}'_{ij}$ & subspace of mismatched signal space of class $i$ that is not associated with the mismatched signal space of class $j$ \\
$\im(\tbU_{ji}' )$ & $ \ti{r}'_{ji}$ & subspace of mismatched signal space of class $j$ that is not associated with the mismatched signal space of class $i$ \\
$\im(\bW_{ij} )$ & $ s^W_{ij}$ & subspace of signal space of class $i$ that is orthogonal to $\im(\tbU_{ij}' )$  \\
$\im(\bV_{ij} )$ & $ s^V_{ij}$ & subspace of signal space of class $i$ that is not orthogonal to $\im(\tbU_{ij}' )$, i.e. it complements $\im(\bW_{ij} )$ in $\im (\bU_i)$
\end{tabular}
\end{table*}

\begin{table}[t]
\caption{Relationships between the quantities used in the analysis} \label{tab:relationships}
\centering
\begin{tabular}{p{3in}}
\begin{enumerate}
\item $ \im(\tbU_i) + \im(\tbU_j) = \im(\tbU_{ij}') + \im(\tbU_{ij}^\cap) + \im(\tbU_{ji}')$  
\item $\im(\tbU_{ij}^\cap)  = \im(\tbU_{ji}^\cap)  = \im(\tbU_i) \cap \im(\tbU_j)$
\item $ \im(\tbU_{ij}') = \im(\tbU_i) \cap \im(\tbU_{ij}^\cap)^\perp$ 
\item $ \im(\tbU_{ji}') = \im(\tbU_j) \cap \im(\tbU_{ij}^\cap)^\perp$ 
\item $ \im(\tbU_{ij}')^\perp = \im(\tbU_i)^\perp + \im(\tbU_{ij}^\cap)$ 
\item $ \im(\bU_i) = \im(\bW_{ij}) + \im(\bV_{ij})$ 
\item $ \im(\bW_{ij}) = \im(\bU_i) \cap \im (\tbU_{ij}')^\perp$ 
\item $ \im(\bV_{ij}) = \im(\bU_i) \cap \im(\bW_{ij})^\perp $ 
\end{enumerate}
\end{tabular}
\vspace*{-15pt}
\end{table}

\subsubsection{Quantities associated with the geometry of the true signal model or the mismatched signal model} \label{sec:quantities_true}

The signal space corresponding to class $i$ and the mismatched signal space corresponding to class $i$, which are subspaces of $\Rn$, are denoted as $\im(\bSigma_i)$ and $\im(\tbSigma_i)$, respectively. An orthonormal basis for $\im(\bSigma_i)$ is denoted as $\bU_i \in \Rt{N}{r_{i}}$ and an orthonormal basis for $\im(\tbSigma_i)$ is denoted as $\tbU_i \in \Rt{N}{\ti{r}_{i}}$; these quantities follow directly from the truncated eigenvalue decompositions $\bSigma_{i} = \bU_{i} \bLambda_{i} \bU_{i}^{T}$ and $\tbSigma_{i} = \tbU_{i} \tbLambda_{i} \tbU_{i}^{T}$ where $\bLambda_{i} = \diag(\lambda_{1}^{i}, \lambda_{2}^{i}, \ldots, \lambda_{r_{i}}^{i}) \in \Rt{r_{i}}{r_{i}}$ and $\tbLambda_{i} = \diag(\ti{\lambda}_{1}^{i}, \ti{\lambda}_{2}^{i}, \ldots, \ti{\lambda}_{\ti{r}_{i}}^{i}) \in \Rt{\ti{r}_{i}}{\ti{r}_{i}}$  are diagonal matrices containing the positive eigenvalues of $\bSigma_{i}$ and $\tbSigma_{i}$, respectively. Note that $\im(\bSigma_i) = \im(\bU_i)$ and $\im(\tbSigma_i) = \im(\tbU_i)$.

\subsubsection{Quantities associated with the interplay between the geometry of the mismatched signal models} \label{sec:quantities_mis}
We consider quantities that reveal the relationship between the mismatched signal spaces of classes $i$ and $j$. In particular, such quantities follow from the decomposition of the subspace $\im(\tbSigma_i + \tbSigma_j) =  \im(\tbU_i) + \im(\tbU_j)$, which spans the mismatched signal subspaces of classes $i$ and $j$, given by:
	\begin{IEEEeqnarray}{rCl}
		 \im(\tbU_i) + \im(\tbU_j) = \lefteqn{\underbrace{\phantom{\im(\tbU_{ij}') + \im(\tbU_{ij}^\cap)}}_{\im(\tbU_i) = \im(\tbSigma_i)}} \im(\tbU_{ij}') + \overbrace{\im(\tbU_{ij}^\cap) + \im(\tbU_{ji}')}^{\im(\tbU_j) = \im(\tbSigma_j)}   \nonumber 
	\end{IEEEeqnarray}
where 
\begin{itemize}	
	\item $\tbU_{ij}^{\cap} \in \Rt{N}{\ti{r}_{ij}^{\cap}}$ represents an orthonormal basis for the intersection $\im(\tbSigma_{i}) \cap \im(\tbSigma_{j})$ and $\ti{r}_{ij}^{\cap}$ is the dimension of $\im(\tbSigma_{i}) \cap \im(\tbSigma_{j})$. This intersection is associated with class $i$ as well as class $j$;
	\item $\tbU_{ij}' \in \Rt{N}{\ti{r}'_{ij}}$ represents an orthonormal basis for the orthogonal complement of $\im(\tbSigma_{i}) \cap \im(\tbSigma_{j})$ in $\im(\tbSigma_{i}) $ and $\ti{r}'_{ij}$ is the codimension of $\im(\tbSigma_{i}) \cap \im(\tbSigma_{j})$ in $\im(\tbSigma_{i}) $. $\im(\tbU_{ij}')$ can be interpreted as the subspace of the mismatched signal space corresponding to class $i$ that is only associated with class $i$ and not with class $j$;
	
	\item $\tbU_{ji}' \in \Rt{N}{\ti{r}'_{ji}}$ represents an orthonormal basis for the orthogonal complement of $\im(\tbSigma_{i}) \cap \im(\tbSigma_{j})$ in $\im(\tbSigma_{j}) $ and $\ti{r}'_{ji}$ is the codimension of $\im(\tbSigma_{i}) \cap \im(\tbSigma_{j})$ in $\im(\tbSigma_{j}) $. $\im(\tbU_{ij}')$ can be interpreted as the subspace of the mismatched signal space corresponding to class $j$ that is only associated with class $j$ and not with class $i$.
\end{itemize}	
Note that $\tbU_{ij}^{\cap}$ together with $\tbU_{ij}'$ and $\tbU_{ji}'$ complete the basis for $\im(\tbU_{i})$ and $\im(\tbU_{j})$, respectively, i.e. $\im(\tbU_{i}) = \im([\tbU_{ij}^{\cap} \, \tbU_{ij}' ])$  and $\im(\tbU_{j}) = \im([\tbU_{ij}^{\cap} \, \tbU_{ji}' ])$.

\subsubsection{Quantities associated with the interplay between the geometry of the true signal model and the mismatched signal model} \label{sec:quantities_inter} We also consider quantities that capture the interaction between the signal space corresponding to class $i$ and the mismatched signal spaces of classes $i$ and $j$. Such quantities  are given by the decomposition of $\im(\bSigma_i) = \im(\bU_i)$ given by:
\begin{IEEEeqnarray}{rCl}
	\im(\bU_i) = \im(\bW_{ij}) + \im(\bV_{ij}) \nonumber
\end{IEEEeqnarray}
where
\begin{itemize}
	\item $\im(\bW_{ij}) = \im(\bU_i) \cap \im (\tbU_{ij}')^\perp$ and 
	 $\bW_{ij} \in \Rt{N}{s^W_{ij}} $ represents an orthonormal basis for $\im(\bU_{i}) \cap \im (\tbU_{ij}')^\perp$ where $s^W_{ij}  = \dim(\im(\bU_{i}) \cap \im (\tbU_{ij}')^\perp)$. $\im(\bW_{ij})$ can be interpreted as the subspace of signal space corresponding to class $i$ that is orthogonal to $\im(\tbU_{ij}' )$;
	
	\item $\im(\bV_{ij}) = \im(\bU_i) \cap (\im(\bU_i) \cap \im (\tbU_{ij}')^\perp )^\perp$ and $\bV_{ij} \in \Rt{N}{s^V_{ij}} $ represents an orthonormal basis for the orthogonal complement of  $\im(\bU_{i}) \cap \im(\tbU_{ij}')^\perp$ in $\im(\bU_{i})$ where $s^V_{ij} = r_{i} - \dim(\im(\bU_{i}) \cap \im(\tbU_{ij}')^\perp)$; then, $s^V_{ij} = \dim(\im(\bV_{ij})) = \rank(\bV_{ij})$ is the codimension of $\im(\bW_{ij})$ in $\im(\bU_i)$. $\im(\bV_{ij})$ can be interpreted as the subspace of signal space of class $i$ that is not orthogonal to $\im(\tbU_{ij}' )$, i.e. it complements $\im(\bW_{ij} )$ in $\im (\bU_i)$.
\end{itemize}
Note that $\im( [ \bV_{ij} \, \bW_{ij}]) = \im(\bU_{i})$.


 
}
\subsubsection{Principal angles and distance between subspaces} 

Finally, our results will also be expressed via the principal angles between certain subspaces. In particular,
consider a subspace $\mathcal{Y}$ with an orthonormal basis $\bY \in \Rt{N}{y}$, where $y=\dim(\mathcal{Y})$, and a subspace $\mathcal{Z}$ with an orthonormal basis $\bZ \in \Rt{N}{z}$, where $z=\dim(\mathcal{Z})$, and define $k = \min(y,z)$. Then the principal angles $0 \leq \theta_1 \leq \cdots \leq \theta_k \leq \frac{\pi}{2}$ between  $\mathcal{Y}$ and $\mathcal{Z}$  are given by the singular value decomposition (SVD):
\begin{IEEEeqnarray}{c}
	\bY^{T} \bZ = \bH \bD \bJ^{T} \label{eq:svdPA}
\end{IEEEeqnarray}
where $\bH \in \Rt{y}{y}$ and $\bJ \in \Rt{z}{z}$ are orthonormal matrices and $\bD \in \Rt{y}{z}$ is a rectangular diagonal matrix containing the singular values: $1 \geq d_{1} \geq \cdots \geq d_{k} \geq 0$. Each singular value $d_{l}$ corresponds to the cosine of the principal angle $\theta_{l}$ between $\mathcal{Y}$ and $\mathcal{Z}$, i.e., $d_{l} = \cos(\theta_{l})$ \cite[Chapter 8.7]{Golub2012}.

The principal angles are used to define various distances on a Grassmann manifold \cite{Hamm2008}. We will be predominantly using the max correlation distance between two subspaces 
\begin{IEEEeqnarray}{rCl}
	d_{\max}(\mathcal{Y}, \mathcal{Z}) &=& d_{\max}(\bY, \bZ)  = \sin \theta_{1} \label{eq:dmax}
\end{IEEEeqnarray}
which is a function of the smallest principal angle $\theta_{1}$, and the min correlation distance between two subspaces
\begin{IEEEeqnarray}{rCl}
	d_{\min}(\mathcal{Y}, \mathcal{Z}) &=& d_{\min}(\bY, \bZ)= \sin  \theta_{k} \label{eq:dmin}
\end{IEEEeqnarray}
which is a function of the largest principal angle $\theta_{k}$ between the two subspaces. Note that we slightly abuse the notation in the second term of \eqref{eq:dmax} and \eqref{eq:dmin}, as $\bY$ and $\bZ$ are bases for the subspaces, not subspaces.


\review{
\subsubsection{Interpretation} 
It is instructive to cast some insight on the role of these various quantities in the characterization of the performance of the MMAP classifier.

Consider a two-class classification problem that involves distinguishing class $1$ from class $2$ in the low-noise regime (so $\by \approx \bx$). It is clear that the MMAP classifier will associate an observation $\by \in \im(\tbU_{12}')$ with class $1$ and an observation $\by \in \im(\tbU_{21}')$ with class $2$; in turn, the MMAP classifier may associate an observation $\by \in \im(\tbU^{\cap}_{12})$ either with class $1$ or $2$. In general, the observation associated with class $1$ is such that $\by \in \im(\bU_1) =  \im(\bV_{12}) + \im(\bW_{12})$.

The following example demonstrates the classification of $\by | c = 1$ by the MMAP classifier where the covariance matrices are assumed to be diagonal.
\vspace{0.25cm}
\begin{example} \label{th:example_diagonal}
We take the covariance matrices to be
\begin{IEEEeqnarray*}{rCl}
	\bSigma_1 &=& \diag(1,1,1,0)\,, \quad \bSigma_2 = \diag(0,1,1,1)	 \\
	\tbSigma_1 &=& \diag(1,1,0,0)\,, \quad	\tbSigma_2 = \diag(0,1,1,0) \,.	
\end{IEEEeqnarray*}
The relevant quantities (see Table~\ref{tab:quantities}) are given as:
\begin{IEEEeqnarray*}{rll}
	&\bU_1 = [\be_1, \be_2, \be_3]\,, \quad	&\bU_2 = [\be_2, \be_3, \be_4] \\	
	&\tbU_1 = [\be_1, \be_2]\,, \quad &\tbU_2 = [\be_2, \be_3]	\,,	 
\end{IEEEeqnarray*}
and
\begin{IEEEeqnarray*}{rCl}
	\tbU_{12}' = \be_1\,, \quad	\tbU_{12}^\cap = \be_2 \,, \quad \tbU_{21}' = \be_3	\,.	
\end{IEEEeqnarray*}
We also determine $\im(\bW_{12})$ and $\im(\bV_{12})$:
\begin{IEEEeqnarray*}{rCl}
\im(\bW_{12}) &=& \im(\bU_1) \cap \im(\tbU_{12}')^\perp \\
	&=& \im([\be_1, \be_2, \be_3]) \cap \im([\be_2, \be_3, \be_4]) = \im([\be_2, \be_3]) \\
\im(\bV_{12}) &=& \im(\bU_1) \cap \im(\bW_{12})^\perp = \be_1 \,.
\end{IEEEeqnarray*}

Assume now that $\by \in \im(\bV_{12})$ and note that $\im(\bV_{12}) = \im(\tbU_{12}')$. Therefore, $\by \in \im(\bV_{12})$ will be classified as class 1 by the MMAP classifier. In contrast, assume now that $\by \in \im(\bW_{12})$ and note that $\im(\bW_{12})$ contains $\im(\tbU_{21}')$. Therefore $\by \in \im(\bW_{12})$ may be classified as class 2. 

Next, we modify the mismatched model of class 2 as
\begin{IEEEeqnarray*}{rCl}
	\tbSigma_2 &=& \diag(0,1,0,1)\,,
\end{IEEEeqnarray*}
which leads to $\tbU_{21}' = \be_4	$. Note now that $\im(\bW_{12})$ does not contain $\im(\tbU_{21}')$ and $\by \in \im(\bW_{12})$ will not be associated uniquely with class 2 by the MMAP classifier.

It is now clear that the relationship between subspaces $\im(\bW_{12})$ and $\im(\tbU_{21}')$ will play a role in the characterization of conditions for perfect classification in the low-noise regime.
\end{example}
\vspace{0.25cm}
The next example demonstrates the role of principal angles in the conditions for perfect classification in the low-noise regime. 
\vspace{0.25cm}
\begin{example} \label{th:example_angles}
We take the signal space bases as:
\begin{IEEEeqnarray*}{l}
	\bU_{1} = [0, 1]^T \,, \quad	\bU_{2} =  \left[\cos \left(\frac{\pi}{4} \right)\,,  \sin \left(\frac{\pi}{4} \right) \right]^{T} \\	
	\tbU_{1} = \left[\cos \left(\frac{5\pi}{6} \right),  \sin \left(\frac{5\pi}{6}\right) \right]^{T} \,, \quad 	\tbU_{2} =  \bU_{2} \,.
\end{IEEEeqnarray*}
The relevant quantities (see Table~\ref{tab:quantities}) are given as:
\begin{IEEEeqnarray*}{rCl}
	\tbU_{12}' = \tbU_1 \,, \quad \tbU_{12}^\cap = \{0\}\,, \quad \tbU_{21}' = \tbU_2
\end{IEEEeqnarray*}
and $\bW_{12} = \{0\}$, $\bV_{12} = \bU_1$.
The geometry of the signals and decision regions is presented in Fig. \ref{fig:angles} (a). Note now that $\by|c=1  \in \im (\bU_1)$ can potentially be associated to the correct class $1$ depending on the distance (computed according to an appropriate metric) between $\im(\bV_{12})$ and $\im(\tbU_{12}')$ and the distance between $\im(\bV_{12})$ and $\im(\tbU_{21}')$. In particular, the angle between $\im(\bV_{12})$ and $\im(\tbU_{12}')$ is greater than the angle between $\im(\bV_{12})$ and $\im(\tbU_{21}')$, which leads to misclassification of signals from \mbox{class 1}. On the other hand, if we take 
\begin{IEEEeqnarray*}{rCl}
	\tbU_{1} &=& \left[\cos \left(\frac{4\pi}{6} \right),  \sin \left(\frac{4\pi}{6}\right) \right]^{T}
\end{IEEEeqnarray*}
the angle between $\im(\bV_{12})$ and $\im(\tbU_{12}')$ is smaller than the angle between $\im(\bV_{12})$ and $\im(\tbU_{21}')$, which leads to perfect classification of signals from class 1 in the low-noise regime. This case is presented in Fig. \ref{fig:angles} (b).
\end{example}
\vspace{0.25cm}

\begin{figure}[t]
	\centering
	\subfigure[\mbox{Example of wrong classification with the MMAP classifier}]{
		\def\XYmax{2}
		\def\angleSigO{90}
		\def\angleSigM{150} 
		\def\angleSigT{45}
		\def\boundTanA{-0.1324}
		\def\boundTanB{-0.1324}
		\def\boundAngleC{0}
		\def\boundAngleD{0}
		
		\begin{tikzpicture}[font =\footnotesize]
		\begin{axis}[width=0.3\textwidth,color=black,axis lines=box,xlabel = {$\mathbf{e}_{1}$},ylabel = {$\mathbf{e}_{2}$},		    
		    xmin=-\XYmax, xmax=\XYmax,
		    ymin=-\XYmax, ymax=\XYmax,
		    xtick=\empty, ytick=\empty
		]
			\addplot[fill=blue, opacity = 0.3,draw=none] coordinates 
			{(0,0) (-\XYmax, \boundTanA*\XYmax) (-\XYmax,\XYmax) (\XYmax*\boundTanB,\XYmax)};
			\addplot[fill=red, opacity = 0.3,draw=none] coordinates 
			{(0,0)  (\XYmax*\boundTanB,\XYmax) (\XYmax,\XYmax) (\XYmax, -\boundTanA*\XYmax)};
			\addplot[fill=blue, opacity = 0.3,draw=none] coordinates 
			{(0,0)  (\XYmax,  -\boundTanA * \XYmax) (\XYmax, -\XYmax) (-\XYmax*\boundTanB, -\XYmax)};		
			\addplot[fill=red, opacity = 0.3,draw=none] coordinates 
			{(0,0)  (-\XYmax*\boundTanB, -\XYmax) (-\XYmax, -\XYmax) (-\XYmax, \boundTanA*\XYmax)};		
			\addplot [domain=-\XYmax:\XYmax, samples=2, dashed] {x*tan(\angleSigO)};
			\addplot [domain=-\XYmax:\XYmax, samples=2, solid,line width=3, color = blue] coordinates {(0, -\XYmax) (0, \XYmax)};
			\addplot [domain=-\XYmax:\XYmax, samples=2, solid,line width=3, color = red] {x*tan(\angleSigT)};
			\addplot [domain=-\XYmax:\XYmax, samples=2, dashed,line width=3, color = blue] {x*tan(\angleSigM)};		
		\end{axis}
		
		\draw node at (2.6,2.9) [color=black,rectangle,  style={align=center}]  {$\text{im}(\mathbf{U}_{1})$};
		\draw node at (3.3,2) [color=black,rectangle,  style={align=center}]  {$\text{im}({\mathbf{U}}_{2})$};
		\draw node at (0.7,1.7) [color=black,rectangle,  style={align=center}]  {$\text{im}(\tilde{\mathbf{U}}_{1})$};

		\end{tikzpicture}				
	} 
	\subfigure[\mbox{Example of correct classification with the MMAP classifier}]{
		\def\XYmax{2}
		\def\angleSigO{90}
		\def\angleSigM{120} 
		\def\angleSigT{45}
		\def\boundTanA{0.1324}
		\def\boundTanB{0.1324}
		\def\boundAngleC{0}
		\def\boundAngleD{0}
		\begin{tikzpicture}[font =\footnotesize]
		\begin{axis}[width=0.3\textwidth,color=black, axis lines=box,xlabel = {$\mathbf{e}_{1}$},ylabel = {$\mathbf{e}_{2}$},		    
		    xmin=-\XYmax, xmax=\XYmax,
		    ymin=-\XYmax, ymax=\XYmax,
		    xtick=\empty, ytick=\empty
		]
			\addplot[fill=blue, opacity = 0.3,draw=none] coordinates 
			{(0,0) (-\XYmax, \boundTanA*\XYmax) (-\XYmax,\XYmax) (\XYmax*\boundTanB,\XYmax)};

			\addplot[fill=red, opacity = 0.3,draw=none] coordinates 
			{(0,0)  (\XYmax*\boundTanB,\XYmax) (\XYmax,\XYmax) (\XYmax, -\boundTanA*\XYmax)};

			\addplot[fill=blue, opacity = 0.3,draw=none] coordinates 
			{(0,0)  (\XYmax,  -\boundTanA * \XYmax) (\XYmax, -\XYmax) (-\XYmax*\boundTanB, -\XYmax)};
		
			\addplot[fill=red, opacity = 0.3,draw=none] coordinates 
			{(0,0)  (-\XYmax*\boundTanB, -\XYmax) (-\XYmax, -\XYmax) (-\XYmax, \boundTanA*\XYmax)};
		
			\addplot [domain=-\XYmax:\XYmax, samples=2, solid,line width=3, color = blue] coordinates {(0, -\XYmax) (0, \XYmax)};
			\addplot [domain=-\XYmax:\XYmax, samples=2, solid,line width=3, color = red] {x*tan(\angleSigT)};
			\addplot [domain=-\XYmax:\XYmax, samples=2, dashed,line width=3, color = blue] {x*tan(\angleSigM)};		
		\end{axis}
		
				\draw node at (2.6,2.9) [color=black,rectangle,  style={align=center}]  {$\text{im}(\mathbf{U}_{1})$};
		\draw node at (3.3,2) [color=black,rectangle,  style={align=center}]  {$\text{im}({\mathbf{U}}_{2})$};
		\draw node at (0.8,2.1) [color=black,rectangle,  style={align=center}]  {$\text{im}(\tilde{\mathbf{U}}_{1})$};		
		\end{tikzpicture}

	} 	
	\vspace*{-7pt}
	\caption{The two plots illustrate the decision regions associated with the 2-class MMAP classifier for different values of $\bU_{1}$, $\bU_{2}$, $\tbU_{1}$ and $\tbU_{2}$ in the limit $\sn \to 0$. Transparent blue and red regions indicate the decision region where MMAP outputs class labels 1 and 2, respectively. Blue line represent the signal subspace $\im(\bU_{1})$ and red line  represent the signal subspace $\im(\bU_{2})$. Dashed blue line represents the mismatched signal subspace $\im(\tbU_1)$. The subspace bases are given in Example 2.}  \label{fig:angles}
	\vspace*{-10pt}
\end{figure}

}

The ensuing analysis shows how these various quantities -- which are readily computed from the underlying geometry of the true subspaces and the mismatched ones -- can be used as a proxy to define sufficient conditions for perfect classification in the low-noise regime. In particular, these quantities bypass the need to compute the decision regions associated with the MMAP classifier in order to quantify the performance.

\section{Conditions for Reliable Classification} \label{sec:conditions}

We now consider (sufficient) conditions for reliable classification in the low-noise regime. We derive these conditions directly from a low-noise expansion of an upper bound to the error probability associated with the MMAP classifier.

The following upper bound to the probability of error associated with a MMAP classifier will play a key role in the analysis.
\vspace{0.15cm}
\begin{theorem}
Set $\alpha_{ij} > 0 \, \forall (i, j) \,, i \neq j $. Set
\begin{IEEEeqnarray}{rcl}
\bSigma_{ij} =  (\bSigma_{i} + \sn \bI)^{-1}  &+& \alpha_{ij} (\tbSigma_{j} + \sn \bI)^{-1}  \nonumber \\
		  &-& \alpha_{ij} (\tbSigma_{i} + \sn \bI)^{-1} \,.
\end{IEEEeqnarray}
Then the error probability associated with the MMAP classifier in \eqref{eq:MMAPerror} can be bounded as follows:
\begin{itemize}
	\item If $\bSigma_{ij} \succ \bZero \, \forall (i,j)$  with $i \neq j $, then
	\begin{IEEEeqnarray}{c}
		P(e) \leq \bar{P}(e) = \sum_{i=1}^{C} p_{i} \cdot \left( \sum_{j = 1, j \neq i}^{C} \bar{P}(e_{ij}) \right) \label{eq:MMAPerrorUB}
	\end{IEEEeqnarray}
	where
		\begin{IEEEeqnarray}{rCl}
			\bar{P}(e_{ij}) &=&  \left(\frac{\ti{p}_j }{\ti{p}_i} \sqrt{\frac{|\tbSigma_{i} + \sn \bI|}{|\tbSigma_{j} + \sn \bI|}} \right)^{\alpha_{ij}} \cdot (|\bSigma_i + \sn \bI| |\pmmSigma{i}{j}|)^{-\oneo{2}} \,.  \nonumber \\ \label{eq:MMAPerrorUBij}
		\end{IEEEeqnarray}
		
	\item If $\exists (i,j)$ with $i \neq j: \bSigma_{ij} \not \succ \bZero$ then $P(e) \leq \bar{P}(e) = 1 $.
\end{itemize}
\vspace{0.15cm} 
\begin{IEEEproof}
	The proof appears in Appendix.
\end{IEEEproof}
\end{theorem}	
\vspace{0.15cm}
\review{
This upper bound to the error probability of the MMAP classifier can capture the fact that the error probability may tend to zero as the noise power approaches zero, depending on the relation between the true signal parameters and the mismatched ones. In particular, the upper bound to the misclassification probability of class $i$ is expressed as a function of the covariance matrix of class $i$, the mismatched covariance matrix of class $i$ and the mismatched covariance matrices of classes $j \neq i$. In contrast, the bound proposed in \cite{Schluter2013}  expresses the upper bound to the error probability as a function of the sum of $f$-divergences between the true and the mismatched distributions of class $i$, for all classes $i$. Therefore, it does not capture the interplay between mismatched models of different classes. In addition, when specialized to the proposed signal model, the bound in \cite{Schluter2013} always predicts the presence of an error floor (see Section \ref{sec:experiments}).} 

The following Theorem presents a low-noise expansion of the upper bound to the error probability of the MMAP classifier.

\vspace{0.15cm}
\begin{theorem}
	The upper bound to the error probability of the MMAP classifier in \eqref{eq:MMAPerrorUB} can be expanded as follows:
	\begin{itemize}
		\item Assume that $\forall (i,j)\,, i \neq j$, the following conditions hold:
			\begin{IEEEeqnarray}{c}
				\im(\bW_{ij}) \subseteq \im(\tbU_{ji}')^\perp \label{eq:nec} \,,
			\end{IEEEeqnarray}
			\begin{IEEEeqnarray}{c}
				\bV_{ij}^{T} (\tbU_{ij}' (\tbU_{ij}')^{T} - \tbU_{ji}' (\tbU_{ji}')^{T}) \bV_{ij} \succ \bZero \text{ or } s^V_{ij} = 0\,, \label{eq:suff} \nonumber \\*
			\end{IEEEeqnarray}
			and take	 $d = \min_{(i \neq j)} d_{ij}$, 		where
			\begin{IEEEeqnarray}{c}
				d_{ij} = \oneo{2} \left ( s^V_{ij} + \alpha_{ij} (\ti{r}_{j} - \ti{r}_{i})\right)\,, \label{eq:Th2_dij}
			\end{IEEEeqnarray}
	and $\alpha_{ij} \in (0, \alpha_{ij}^{0})$ where the value of $\alpha_{ij}^{0}>0$ is given in the Appendix.		
			Then
			\begin{itemize}
			\item If $d \leq 0$: 
				\begin{IEEEeqnarray}{c}
					\bar{P} (e) = \mathcal{O} (1), \qquad \sigma^2 \to 0 \,.
				\end{IEEEeqnarray}
			\item If $d > 0$:	
				\begin{IEEEeqnarray}{rCl}
					\bar{P}(e) &=& A\cdot \left(\sn \right)^{d} 
					 + o\left( \left( \sn \right)^{d} \right)\,,  \,\,\,\,\, \sn \to 0\,,
					 \label{eq:Peexpansion}
				\end{IEEEeqnarray}
			\end{itemize}				
			where $A > 0$.

	\item Assume $\exists (i,j)$, $i \neq j$, such that conditions \eqref{eq:nec} or \eqref{eq:suff} do not hold. Then
				\begin{IEEEeqnarray}{c}
					\bar{P} (e) = \mathcal{O} (1), \qquad \sigma^2 \to 0 \,.
				\end{IEEEeqnarray}
	\end{itemize}
	
\vspace{0.15cm} 
\begin{IEEEproof}
	The proof appears in Appendix.
\end{IEEEproof}
\end{theorem}
\vspace{0.15cm}
	
The expansion of the upper bound to the error probability embodied in Theorem 2 provides a set of conditions, which are a function of the geometry of the true signal model, the geometry of the mismatched signal model, and the interaction of the geometries, that enable us to understand whether or not the upper bound to the error probability may exhibit an error floor.  In particular, in view of the fact that we use the union bound in order to bound the error probability of a multi-class problem in terms of the error probabilities of two-class problems, these conditions have to hold for every pair of class labels $(i,j)$, $i \neq j$. We can note that:
\begin{itemize}

\item \review{The upper bound to the probability of error exhibits an error floor if either \eqref{eq:nec} or \eqref{eq:suff} are not satisfied for some pair $(i,j)$, $i\neq j$. The interpretation of condition \eqref{eq:nec} is straightforward by noting that the subspace $\im(\bW_{ij})$ contains vectors of class $i$ that are orthogonal to the subspace $\im(\bU_{ij}')$, which is the subspace uniquely associated with class $i$. Then, condition \eqref{eq:nec} states that such vectors must also be orthogonal to the mismatched subspace uniquely associated with class $j$, i.e. $\im(\bU_{ji}')$.


The interpretation of condition \eqref{eq:suff} is obtained by reformulating the expression as:
\begin{IEEEeqnarray}{c}
	\bV_{ij}^{T} (\tbU_{ij}' (\tbU_{ij}')^{T} - \tbU_{ji}' (\tbU_{ji}')^{T}) \bV_{ij} \succ \bZero  \label{eq:sufflhs}
	\\ \iff \nonumber \\ 
	\bx^T \tbU_{ij}' (\tbU_{ij}')^{T} \bx > \bx^T \tbU_{ji}' (\tbU_{ji}')^{T} \bx \quad \forall \bx \in \im(\bV_{ij}) \nonumber \\ \iff  \nonumber \\
	\| (\tbU_{ij}')^{T} \bx \|_2 > 	\| (\tbU_{ji}')^{T} \bx \|_2 \quad \forall \bx \in \im(\bV_{ij})  \,. \nonumber
\end{IEEEeqnarray}
Note that $\| (\tbU_{ij}')^{T} \bx \|_2 = \| \tbU_{ij}(\tbU_{ij}')^{T} \bx \|_2$ is the norm of the projection of $\bx$ onto $\im(\tbU_{ij}')$. Therefore, \eqref{eq:suff} requires that the norm  of vectors in $\im(\bV_{ij})$, which are associated with class $i$, projected onto $\im(\tbU_{ij}')$, which is also associated with class $i$, is greater than  the norm  of vectors in $\im(\bV_{ij})$ projected onto $\im(\tbU_{ji}')$, which is associated with \mbox{class $j$}.

Equation \eqref{eq:sufflhs} is also implied by 
\begin{IEEEeqnarray}{c}
	d_{\min}(\bV_{ij}, \tbU_{ij}') < d_{\max}(\bV_{ij}, \tbU_{ji}') \label{eq:distineq}
\end{IEEEeqnarray}	
which requires that the largest principal angle between $\im(\bV_{ij})$ and $\im(\tbU_{ij}')$ is smaller than the smallest principal angle between $\im(\bV_{ij})$ and $\im(\tbU_{ji}')$.\footnote{The detailed derivation of this statement is reported in Appendix.} Demonstration of this condition is provided by Example \ref{th:example_angles}  in Section \ref{sec:geom}.}

\item On the other hand, the upper bound to the probability of error does not exhibit an error floor if conditions \eqref{eq:nec} and \eqref{eq:suff} are satisfied for all pairs $(i,j)$, $i\neq j$ and $d > 0$. In particular, necessary and sufficient conditions for $d>0$ depend on the dimension of the various subspaces and their relation, i.e. $s^V_{ij}> 0$ for all pairs $(i,j)$ such that $\ti{r}_j - \ti{r}_i \leq 0$ is necessary and sufficient for $d>0$. For example, if the rank of all covariance matrices associated to the mismatched model is the same, i.e., if $\ti{r}_{i} = \ti{r}$, for $i = 1, \ldots,  C$, then $s^V_{ij} > 0$, $\forall (i,j)$, $i \neq j$ is necessary and sufficient for $d>0$. Note that  a positive value for $s^V_{ij}$ indicates that there is at least one vector in $\im(\bU_{i})$ that is not contained in $\im(\tbU_{ij}')^\perp$, or equivalently, there exists at least one vector in $\im(\bU_{i})$ that has a non-zero projection onto $\im(\tbU_{ij}')$, therefore leading to reliable classification of signals from class $i$. 

\item Note that parameters $\alpha_{ij}$ do not play a role in the characterization of the necessary and sufficient  conditions for $d>0$. In fact, the conditions for $d_{ij} >0$ do not depend on a particular value of $\alpha_{ij}$, provided that $\alpha_{ij} \in (0, \oneo{|\ti{r}_{j} - \ti{r}_{i} |})$.

\item Note also that the value of $d$ represents a measure of robustness against noise in the low-noise regime, as it determines the speed at which the upper bound of the error probability decays with $1/\sigma^2$. In particular, higher values of $d$ will represent higher robustness against noise, in the low-noise regime. For example, on assuming $\tilde{r}_i=\tilde{r}$ for $i=1,\ldots,C$, we observe that larger values of $s^V_{ij}$ correspond to larger values of $d$. Therefore, as expected, higher levels of robustness are obtained when the overlap between $\im(\bU_i)$ and $\im(\tbU_{ij}')^\perp$, i.e. dimension of $\im(\bW_{ij})$, is reduced. 

\review{
We also discuss how the value of $d_{ij}$ in equation \eqref{eq:Th2_dij} relates to the value of $d_{ij}$ for the non-mismatched case.\footnote{Note that our comparison involves upper bounds on the error probabilities rather than the actual error probabilities.} In particular, we  assume that $r_i = r_j = \ti{r}_i = \ti{r}_j$ and that the true and the mismatched covariance matrices are diagonal. Then for the non-mismatched case 
	\begin{IEEEeqnarray*}{rCl}
		d_{ij} = \oneo{2} (r_i - \dim(\im(\bU_i) \cap  \im(\bU_j)) )
	\end{IEEEeqnarray*}
	and for the mismatched case
	\begin{IEEEeqnarray*}{rCl}
		d_{ij} =& \oneo{2} ( r_i& - \dim(\im(\bU_i) \cap  \im(\tbU_j))   \\
		& &-		\dim(\im(\bU_i) \cap  \ke(\tbU_i) \cap \ke(\tbU_j))) \,.
	\end{IEEEeqnarray*}	
	Therefore, in the non-mismatched case $d_{ij}$ is at most $r_i$ and it decreases as the dimension of the intersection of the signal spaces of classes $i$ and $j$ increases. In the mismatched case $d_{ij}$ is also at most $r_i$, but it decreases as the dimension of the intersection of the signal space of class $i$ and the mismatched signal space of class $j$ increases, and as the dimension of the intersection of the signal space of class $i$ and the noise subspace of the mismatched classifier, i.e $\ke(\tbU_i) \cap \ke(\tbU_j)$, increases. It can also be easily verified that the value of $d$ for a non-mismatched 2 class problem obtained in \cite{Reboredo2014} matches the value of $d$ derived via the proposed bound. Note that the bound analyzed in \cite{Reboredo2014} is different than the bound proposed in this paper and it is only valid for non-mismatched models.
}

\item The constant $A$ in \eqref{eq:Peexpansion} distinguishes the upper bounds for different mismatched models with a constant $d$, in the low-noise regime, and is determined as the ratio of volumes of subspaces associated with true and mismatched signal subspaces and their interaction. See Appendix for the detailed expression.

\end{itemize}

Theorem 2 therefore leads immediately to sufficient conditions for reliable classification in the low-noise regime.
\vspace{0.15cm}
\begin{corollary} 
If 
\begin{IEEEeqnarray}{c}
	\im(\bW_{ij}) \subseteq \im(\tbU_{ji}')^\perp \, \forall (i,j),  i \neq j  \,, \\
	\bV_{ij}^{T} (\tbU_{ij}' (\tbU_{ij}')^{T} - \tbU_{ji}' (\tbU_{ji}')^{T}) \bV_{ij} \succ \bZero \, \forall (i,j),  i \neq j 
\end{IEEEeqnarray}
and $s^V_{ij} > 0 \, \forall (i,j)$ such that $\ti{r}_j - \ti{r}_i \leq 0$, then \mbox{$\lim_{\sigma^2 \to 0} P(e) = 0$.}
	\vspace{0.15cm}
	\begin{IEEEproof}
		This follows directly from Theorem 1, since $\lim_{\sn \to 0}\bar{P} (e) = 0 \implies \lim_{\sn \to 0} P(e) = 0$.
	\end{IEEEproof}
\end{corollary}
\vspace{0.15cm}

\begin{corollary} \label{th:pt}
If
\begin{IEEEeqnarray}{c}
	d_{\min}(\bU_{i}, \tbU_{i}) <  d_{\max}(\bU_{i}, \tbU_{j}) \text{ and } s^V_{ij} > 0 \, \forall (i,j),  i \neq j  \IEEEeqnarraynumspace
 \* \label{eq:suffdist}
\end{IEEEeqnarray}
then $\lim_{\sn \to 0} P(e) = 0$.
	\vspace{0.15cm}
	\begin{IEEEproof}
		The proof appears in Appendix.
	\end{IEEEproof}
\end{corollary}
\vspace{0.15cm}

Note that the conditions in Corollary 2 are implied by (hence are weaker) the conditions in Corollary 1.

The conditions for reliable classification are particularly simple for the scenario where true and mismatched covariance matrices are diagonal.
\vspace{0.15cm}
\begin{corollary} 
Assume $\bSigma_{i}$, $\tbSigma_{i}$, $i=1,\ldots,C$ are diagonal. If 
\begin{IEEEeqnarray}{c}
	\im(\bW_{ij}) \subseteq \im(\tbU_{ji}')^\perp \, \forall (i,j),  i \neq j 
	\label{eq:suffdiag}
\end{IEEEeqnarray}
and $s^V_{ij} > 0 \, \forall (i,j)$ such that $\ti{r}_j - \ti{r}_i \leq 0$, then
\mbox{$\lim_{\sn \to 0} P(e) = 0$}.
	\vspace{0.15cm}
	\begin{IEEEproof}
		The proof appears in Appendix.
	\end{IEEEproof}
\end{corollary}
\vspace{0.15cm}
Note that in diagonal case the sufficient conditions for perfect classification simplify only to inclusion of subspaces. Recall the Example 1 where we demonstrate that the signals in $\im(\bW_{ij})$ may be associated with class $i$ or with class $j$. Condition \eqref{eq:suffdiag} formalizes the intuition that the signals in $\im(\bW_{ij})$ must be orthogonal to $\im(\tbU_{ji}')$, which is uniquely associated with \mbox{class $j$.}

We finally illustrate how our conditions cast insight onto the impact of mismatch for a two-class case where the mismatched subspaces are a rotated version of the true signal subspaces.

\vspace{0.15cm}
\begin{example} \label{th:example_perturbation}
Consider a two-class classification problem where $\bx|c=1 \sim \N(\bZero,\bU_1 \bU_1^T)$ and $\bx|c=2 \sim \N(\bZero,\bU_2 \bU_2^T)$ and
	\begin{IEEEeqnarray}{rcl}
		\tbU_{1} = \bQ_{1} \bU_{1} \,, \quad	\tbU_{2} = \bQ_{2} \bU_{2}		 \,,				\label{eq:rot}
	\end{IEEEeqnarray}
	where $\bQ_{1} \in \Rnn$ and $\bQ_{2} \in \Rnn$ are orthogonal matrices, and $s_{12}, s_{21} > 0$.\footnote{This condition insures that the mismatched subspaces are not completely orthogonal to the signal subspaces.} By defining 
	\begin{IEEEeqnarray}{c}
		\epsilon_{1} = \| \bI - \bQ_{1} \|_{2} \,, \quad 	\epsilon_{2} = \| \bI - \bQ_{2} \|_{2} \\
		\delta_{12} = \max_{l} \cos \theta_{l}^{12} = \sqrt{1- d_{\min}^{2}(\bU_{1}, \bU_{2})}�,
	\end{IEEEeqnarray}
	it follows that
		\begin{IEEEeqnarray}{c}
			1 - \delta_{12} >  \epsilon_{1} + \epsilon_{2} \implies \lim_{\sn \to 0} P(e) = 0\,. \label{eq:suffpert}
	\end{IEEEeqnarray}	
	The proof is in the Appendix.
\end{example}
\vspace{0.15cm}

This example provides sufficient conditions for reliable classification in the low-noise regime by relating the degree of mismatch -- measured in terms of the spectral norm of the matrix $\bI-\bQ_i$, $i=1,2$ -- to the minimum principal angle between subspaces. It states that the larger the minimum principal angle between the spaces spanned by signals of class 1 and class 2, i.e. the larger $1-\delta_{12}$, the more robust is the classifier against mismatch, where the level of mismatch is measured by $\epsilon_{1} + \epsilon_{2}$. The maximum robustness against mismatch is obtained when $\delta_{12} = 0$, which means that signals from class 1 and class 2 are orthogonal.

This example also provides a rationale for state-of-the-art feature extraction mechanisms where the signal classes are transformed via a linear operator $\bPhi$  prior to classification. In particular, assume that $\bSigma_{1}$ and $\bSigma_{2}$ correspond to the covariances of signals in class 1 and 2 after the transformation $\bPhi$: the example suggests that the operator $\bPhi$ should transform the signal covariances so that $\delta_{12}$ is small (i.e. so that the signals from class 1 and 2 are close to orthogonal) in order to create robustness against mismatch. Such an approach is considered, for example, in \cite{Qiu2015}, where signals are transformed by a matrix, which promotes large principal angles between the subspaces. Note that the work in \cite{Qiu2015} is not motivated on the basis of robustness against mismatch, but rather on intuitive insight about classification of signals that lie on subspaces.

\section{Numerical Results} \label{sec:experiments}

We now show that our conditions for reliable classification in the low-noise regime are sharp, by revisiting the Examples \ref{th:example_diagonal} and \ref{th:example_angles} presented in Section \ref{sec:geom}. The model parameters and results are summarized in Table~\ref{tab:examples_table}.
\begin{table*}[t]
\caption{Mismatch examples given in Section \ref{sec:geom}.} \label{tab:examples_table}
\vspace*{-6pt}
\centering
\begin{tabular}{c|c|c||c}
	& Model  & $\begin{array}{c} \text{Theory} \\  \lim_{\sn \to 0} \bar{P}(e) \end{array}$ &  $\begin{array}{c} \text{Simulation} \\  \lim_{\sn \to 0} P(e) \end{array}$ \\
\hline
(a) &  $\begin{array}{c}
\bU_{1} = [\be_{1}, \be_2, \be_3] \,,
\bU_{2} = [\be_{2}, \be_{3}, \be_4] \,,
\tbU_{1} = [\be_1, \be_2] \,, \tbU_{2} = [\be_2, \be_3]
\end{array}$  & $>0$ & $>0$
\\ 
(b) &  $\begin{array}{c}
\bU_{1} = [\be_{1}, \be_2, \be_3] \,,
\bU_{2} = [\be_{2}, \be_{3}, \be_4] \,,
\tbU_{1} = [\be_1, \be_2] \,, \tbU_{2} = [\be_2, \be_4]
\end{array}$  & $=0$ & $=0$
\\ 
(c) &  $\begin{array}{c}
\bU_{1} = [0,1]^T \,, 
\bU_{2} = [ \cos \left(\frac{\pi}{4} \right), \sin \left(\frac{\pi}{4} \right)] \,,
\tbU_{1} =[ \cos \left(\frac{5\pi}{6} \right), \sin \left(\frac{5\pi}{6} \right)] \,,
\tbU_{2} = \bU_2
\end{array}$  & $>0$ & $>0$
\\ 
(d) &  $\begin{array}{c}
\bU_{1} = [0,1]^T \,, 
\bU_{2} = [ \cos \left(\frac{\pi}{4} \right), \sin \left(\frac{\pi}{4} \right)] \,,
\tbU_{1} =[ \cos \left(\frac{4\pi}{6} \right), \sin \left(\frac{4\pi}{6} \right)] \,,
\tbU_{2} = \bU_2
\end{array}$  & $=0$ & $=0$
\end{tabular}
\end{table*}

\begin{figure*}[!t]
	\centering
	\subfigure[$\lim_{\sn \to 0} {P}(e)  > 0$.]{
		\newcommand{\XMAX}{1e90}
		\renewcommand{\LW}{1.5}
		\newcommand{\YMIN}{0.005}
		\newcommand{\YMAX}{1.5}
		\begin{tikzpicture}[font=\footnotesize]
		\begin{axis} [width=0.25\textwidth,logPe SNR, ymin= \YMIN, ymax=\YMAX, xmin=1e-10, xmax=\XMAX,xmode=log, ymode=log,  legend style={font=\normalsize}, legend pos=south west,     ylabel={$\log_{10}\left(P(e)\right)$},xlabel={$1/\sigma^2 \text{ }[\text{dB}]$}]
			\addplot [Pe, color=black] table [x index=0, y index=1] {figdata/error_bounds1.txt};
			\addplot [Pe, color=red] table [x index=0, y index=3] {figdata/error_bounds1.txt};
			\addplot [Pe, color=orange,dashed] table [x index=0, y index=2] {figdata/error_bounds1.txt};
		\end{axis}
		\end{tikzpicture}	
	}\hspace{-0.5cm} 
	\subfigure[$\lim_{\sn \to 0} {P}(e)  = 0$. ]{
		\newcommand{\XMAX}{1e90}
		\renewcommand{\LW}{1.5}
		\newcommand{\YMIN}{0.005}
		\newcommand{\YMAX}{1.5}
		\begin{tikzpicture}[font=\footnotesize]
		\begin{axis} [width=0.25\textwidth,logPe SNR, ymin= \YMIN, ymax=\YMAX, xmin=1e-10, xmax=\XMAX,xmode=log, ymode=log,  legend style={font=\normalsize}, legend pos=south west,     ylabel={$\log_{10}\left(P(e)\right)$},xlabel={$1/\sigma^2 \text{ }[\text{dB}]$}]
			\addplot [Pe, color=black] table [x index=0, y index=1] {figdata/error_bounds2.txt};
			\addplot [Pe, color=red] table [x index=0, y index=3] {figdata/error_bounds2.txt};
			\addplot [Pe, color=orange,dashed] table [x index=0, y index=2] {figdata/error_bounds2.txt};
		\end{axis}
		\end{tikzpicture}		
		} \hspace{-0.5cm}
	\subfigure[$\lim_{\sn \to 0} {P}(e)  > 0$. ]{
			\newcommand{\XMAX}{1e90}
		\renewcommand{\LW}{1.5}
		\newcommand{\YMIN}{0.005}
		\newcommand{\YMAX}{1.5}
		\begin{tikzpicture}[font=\footnotesize]
		\begin{axis} [width=0.25\textwidth,logPe SNR, ymin= \YMIN, ymax=\YMAX, xmin=1e-10, xmax=\XMAX,xmode=log, ymode=log,  legend style={font=\normalsize}, legend pos=south west,     ylabel={$\log_{10}\left(P(e)\right)$},xlabel={$1/\sigma^2 \text{ }[\text{dB}]$}]
			\addplot [Pe, color=black] table [x index=0, y index=1] {figdata/error_bounds3.txt};
			\addplot [Pe, color=red] table [x index=0, y index=3] {figdata/error_bounds3.txt};
			\addplot [Pe, color=orange,dashed] table [x index=0, y index=2] {figdata/error_bounds3.txt};
		\end{axis}
		\end{tikzpicture}
	} \hspace{-0.5cm}
	\subfigure[$\lim_{\sn \to 0} {P}(e)  = 0$. ]{
		\newcommand{\XMAX}{1e90}
		\renewcommand{\LW}{1.5}
		\newcommand{\YMIN}{0.005}
		\newcommand{\YMAX}{1.5}
		\begin{tikzpicture}[font=\footnotesize]
		\begin{axis} [width=0.25\textwidth,logPe SNR, ymin= \YMIN, ymax=\YMAX, xmin=1e-10, xmax=\XMAX,xmode=log, ymode=log,  legend style={font=\normalsize}, legend pos=south west,     ylabel={$\log_{10}\left(P(e)\right)$},xlabel={$1/\sigma^2 \text{ }[\text{dB}]$}]
			\addplot [Pe, color=black] table [x index=0, y index=1] {figdata/error_bounds4.txt};
			\addplot [Pe, color=red] table [x index=0, y index=3] {figdata/error_bounds4.txt};
			\addplot [Pe, color=orange,dashed] table [x index=0, y index=2] {figdata/error_bounds4.txt};
		\end{axis}
		\end{tikzpicture}
	} 
	\vspace*{-7pt}
	\caption{Simulation results for the examples in Table 1. \review{
In all plots, the black line corresponds to the true error probability $P(e)$ obtained via simulation, the red line corresponds to the proposed upper bound to error probability $\bar{P}(e)$ given in Theorem 1 and the dashed orange line corresponds to the upper bound in \cite{Schluter2013} (with KL-divergence).
}} \label{fig:examples}
	\vspace*{-10pt}
\end{figure*}
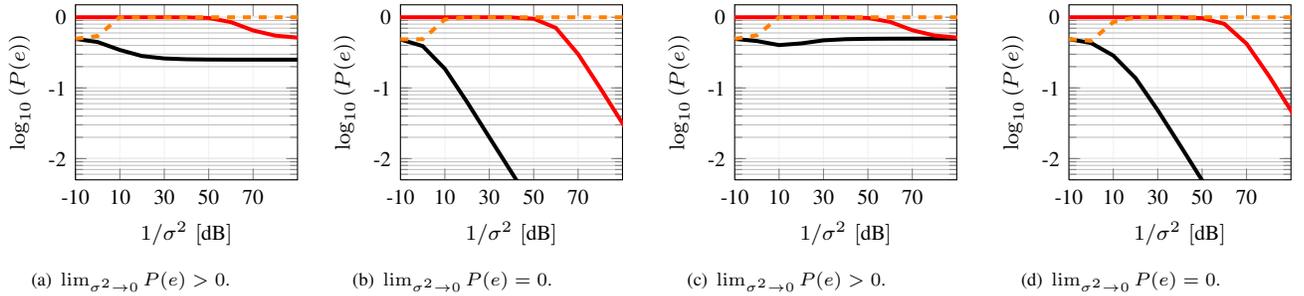

\review{
Fig. \ref{fig:examples} shows the estimated true error probability, which is obtained from simulation\footnote{In our simulations, signals are drawn independently from the true distribution and are classified by the MMAP classifier.}, the upper bound to the error probability given in Theorem 1 and the bound proposed in \cite{Schluter2013} (using the KL-divergence) as a function of $\sn$. Note that the proposed upper bound to the error probability and the derived sufficient conditions give a sharp predictions of an error floor, and also that the bound proposed in \cite{Schluter2013} always exhibits an error floor.
}

In case (a), condition  \eqref{eq:nec} in Theorem 2 is not satisfied for $(i,j) = (1,2)$, i.e. $\im(\bW_{12}) = \im([\be_2, \be_3]) \not\subseteq \im(\tbU_{21}')^\perp = \im(\be_{3})^\perp$, therefore, via Theorem 2 we conclude that the upper bound exhibits an error floor. The results in Fig. \ref{fig:examples} show that in this case the true error probability also exhibits an error floor.
In case (b), conditions \eqref{eq:nec} and \eqref{eq:suff} are satisfied and $d > 0$. Therefore, via Theorem 2, the upper bound to the error probability approaches zero, which also implies that the true error probability approaches zero, in the low-noise regime. 

For cases (c) and (d) the intuition is provided by the Corollary 2, where in the case of the one-dimensional subspaces the concept of principal angles simply reduces to the notion of angle between two lines. In particular, in case (c) the condition \eqref{eq:suffdist} in Corollary 2 is not satisfied for $(i,j) = (1,2)$, and we observe an error floor in the true error probability. On the contrary, in case (d) the conditions \eqref{eq:suffdist} in Corollary 2 are satisfied which immediately implies perfect classification in the low-noise regime.

\begin{figure}[!t]
	\centering
	\newcommand{\XMAX}{1e100}
	\renewcommand{\LW}{1.5}
	\newcommand{\YMIN}{0.0005}
	\newcommand{\YMAX}{1.3}
	\begin{tikzpicture}[font=\footnotesize]
		\begin{axis} [width=0.25\textwidth,logPe SNR, ymin= \YMIN, ymax=\YMAX, xmin=1e-10, xmax=\XMAX, xmode=log, ymode=log,  legend style={font=\normalsize}, legend pos=south west,     ylabel={$\log_{10}\left(P(e)\right)$}, xlabel={$1 / \sigma^2\text{ [dB]}$},xtick={1e-10,1e10,1e30,1e50,1e70,1e90},
    xticklabels={-10,10,30,50,70,90},]
		\addplot [Pe, color=black] table [x index=0, y index=1] {figdata/robust_error.txt};
		\addplot [Pe, color=black,dashed] table [x index=0, y index=1] {figdata/robust_error_bound.txt};
		\addplot [Pe, color=red] table [x index=0, y index=3] {figdata/robust_error.txt};
		\addplot [Pe, color=red,dashed] table [x index=0, y index=3] {figdata/robust_error_bound.txt};
		\addplot [Pe, color=blue] table [x index=0, y index=2] {figdata/robust_error.txt};
		\addplot [Pe, color=blue,dashed] table [x index=0, y index=2] {figdata/robust_error_bound.txt};	
		\end{axis}
	\end{tikzpicture}
	\vspace*{-7pt}
	\caption{Black, blue and red lines correspond to the simulated error probabilities for examples given by \eqref{eq:rob1}, \eqref{eq:rob2} and \eqref{eq:rob3}, respectively. Dashed black, blue and red lines correspond to the upper bound given in Theorem~1 for examples given by \eqref{eq:rob1}, \eqref{eq:rob2} and \eqref{eq:rob3}, respectively.} \label{fig:diagonal}
	\vspace*{-10pt} 
\end{figure}
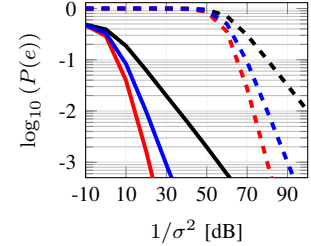

We now explore how different mismatched models affect the value of $d$. Consider the following 2-class example in $\Ro{6}$ with orthonormal basis vectors $\be_{i}$, $i = 1,\ldots, 6$, where the signal spaces are:
\begin{IEEEeqnarray}{c}
	\bU_{1} = [\be_{1}, \be_{2}, \be_{3}] \,, \quad 	\bU_{2} = [\be_{4}, \be_{5}, \be_{6}] 
\end{IEEEeqnarray}
and various mismatched signal spaces are:
\begin{IEEEeqnarray}{lCl}
	\tbU_{1} = [\be_{1}]\,,& \quad 	&\tbU_{2} = [\be_{4}]  \label{eq:rob1}\\
	\tbU_{1} = [\be_{1}, \be_{2}] \,,& 	&\tbU_{2} = [\be_{4}, \be_{5}]  \label{eq:rob2}\\
	\tbU_{1} = \bU_{1} \,, & &	\tbU_{2} = \bU_{2} \,.  \label{eq:rob3}
\end{IEEEeqnarray}
It is straightforward to verify that the sufficient conditions for perfect classification given by Theorem 2 hold for all three pairs of mismatch models \eqref{eq:rob1}, \eqref{eq:rob2} and \eqref{eq:rob3}. Furthermore, one can also determine the values of $d$ as $0.5$, $1$ and $1.5$, where values of $d$ do not depend on $\alpha_{ij}$, for the mismatched models given by \eqref{eq:rob1}, \eqref{eq:rob2} and \eqref{eq:rob3}, respectively. As observed in Section \ref{sec:conditions}, a higher value of $d$ implies a higher robustness to noise. Simulation results of the true error probability and the values of the upper bounds as given in Theorem~1 are plotted in Fig. \ref{fig:diagonal}. One can observe that increasing values of $d$ (associated with the upper bound to the error probability) correspond to steeper decrease of the true error probability as $\sn \to 0$. Moreover, the values of $d$ obtained via the upper bound match the values of $d$ obtained from the simulation of the true error probability for all the examples \eqref{eq:rob1}-\eqref{eq:rob3}.

\section{Applications} \label{sec:applications}

We finally show how theory can also capture the impact of mismatch on classification performance in applications involving real world data. We consider a motion segmentation application, where the goal is to segment a video in multiple rigidly moving objects, and a hand-written digit classification application. In both tasks we concentrate on a supervised learning approach, in which we are given a number of labeled samples, which are used to estimate the model (training set) and a number of unlabeled samples that we want to classify (testing set). Our aim is to determine the minimum size of the training set needed to guarantee reliable classification of the testing set.

\subsection{Datasets}

For the motions segmentation task we use the Hopkins 155 dataset \cite{Tron2007}, which  consists of video sequences with 2 or 3 motions in each video. The motion segmentation problem is usually solved by extracting feature points from the video and tracking their position over different frames. In more details, in this application, observation vectors $\by$ are obtained by stacking the coordinate values associated to a given feature point corresponding to different frames, and the objective of motion segmentation is that of classifying each feature point as belonging to one of the moving objects in the video \cite{vidalTutorial}.

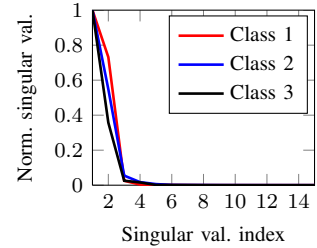
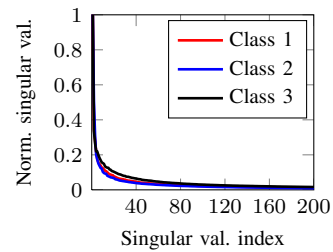
\begin{figure}[t]
	\centering
	\subfigure[Hopkins dataset]{
		\renewcommand{\LW}{1}
		\begin{tikzpicture}[font=\footnotesize]
		    \begin{axis}[width=0.25\textwidth,color=black,ylabel={Norm. singular val.},xlabel= {Singular val. index},
			        xmin = 1, xmax = 15,xtick={2,4,6,8,10,12,14},ymin = 0, ymax = 1,
			        	legend pos = north east,
			]  
		    \addplot[color = red, mark=,line width=\LW] table [x index=0, y index=1] {figdata/hopkins_singular_values.txt};
         	\addlegendentry{Class 1};    
        		\addplot[color = blue, mark=,line width=\LW] table [x index=0, y index=2] {figdata/hopkins_singular_values.txt};
         	\addlegendentry{Class 2};        
        		\addplot[color = black, mark=,line width=\LW] table [x index=0, y index=3] {figdata/hopkins_singular_values.txt};    
         	\addlegendentry{Class 3};            
		    \end{axis} 
		\end{tikzpicture}
	} 
	\subfigure[MNIST dataset]{
		\renewcommand{\LW}{1}	
		\begin{tikzpicture}[font=\footnotesize]	
			\begin{axis}[width=0.25\textwidth,color=black,ylabel={Norm. singular val.},    	xlabel= {Singular val. index},
        		xmin = 0,xmax = 200,xtick={40,80,120,160,200},ymin = 0,ymax = 1,
	        	legend pos = north east,
		]  
    
   		 \addplot[color = red, mark=,line width=\LW] table [x index=0, y index=1] {figdata/mnist_singular_values.txt};
         \addlegendentry{Class 1};
        \addplot[color = blue, mark=,line width=\LW] table [x index=0, y index=2] {figdata/mnist_singular_values.txt};
         \addlegendentry{Class 2};        
        \addplot[color = black, mark=,line width=\LW] table [x index=0, y index=3] {figdata/mnist_singular_values.txt};     
         \addlegendentry{Class 3};           
	    \end{axis} 
		\end{tikzpicture}		
	} 
	\caption{Normalized singular values of data matrices corresponding to: (a) motions in the Hopkins dataset and (b) digits in the MNIST dataset. For Hopkins dataset only the first 15 out of 58 singular values are shown. For MNIST dataset only the first 200 out of 784($=28 \times 28$) singular values are shown for the first 3 classes.
} \label{fig:singular_value}
	\vspace*{-10pt}
\end{figure}

\review{
Theoretical results show that the features points trajectories belonging to a given motion lie on approximately 3 dimensional affine space or 4 dimensional \mbox{linear space \cite{vidalTutorial,Tomasi1992,Boult1991}.} We validate that empirically by observing the decay of singular values of the data matrix associated with a given motion, which is shown in Fig. \ref{fig:singular_value} (a). Note that singular values are close to zero for singular value indices that are \mbox{greater than 4.}}

For the experiment we consider a video with 3 motions\footnote{Denoted as ``1RT2RCR'' in the dataset.}, where number of samples of class 1, class 2 and class 3 is 236, 142 and 114, respectively. The rule adopted to pick the video was the maximal possible feature points -- samples -- for each motion. The ranks of the true and the mismatched covariances is always set to 4. We also split the dataset samples randomly into a training set and a testing set, where the training set contains $n_{\max} = 90$ samples per class.

\review{
For the hand-written digit classification task we use the MNIST dataset \cite{LeCun1998}, which consists of $28 \times 28$ grey scale images of hand-written digits between 0 and 9. We obtain observation vectors $\by$ by vectorizing the images. 

The decay of singular values associated with the data matrix of MNIST digits is shown in Fig. \ref{fig:singular_value} (b). Note that the singular values do not approach zero as fast as in the case of the Hopkins dataset. We can argue that the classes in the MNIST dataset are only ``approximately low-rank'', i.e. the covariance matrix associated with the class $i$ can be expressed as $\bSigma_i = \bar{\bSigma}_i + \delta \bI$, where $\bar{\bSigma}_i$ is low-rank and $\delta > 0$ accounts for the deviation from the perfectly low-rank model. In view of the presented signal model this can be interpreted as a classification of signals with low-rank covariance matrix $\bar{\bSigma}_i$ at finite $\sn = \delta$. The sufficient conditions for perfect classification in the case of ``approximately low-rank'' model will now predict what number of training samples is required to achieve the best possible error rate for the given classification problem. 

The ranks of the true and the mismatched covariances is always set to 20 in the experiments. Such rank leads to capturing approximately 90 \% of the energy of the signals. The split into training and testing set is provided by the MNIST dataset, where the training set contains approximately $n_{max}= 5000$ samples per class. 
}

\subsection{Methodology}

We obtain the class-conditioned covariance matrices by retaining only the first $r$ principal components of the estimated covariances  obtained via the maximum likelihood (ML) estimator\footnote{Note that this is equivalent to computing the empirical covariance matrix.}  for each class. The covariance matrix associated with the ``true model'' of class $i$ is obtained by estimating the covariance matrix on all available data samples of class $i$, and the covariance matrices associated with the ``mismatched model'' of class $i$ are obtained by estimating the covariance matrix on $n_i$ data samples of \mbox{class $i$.}

Results are produced as follows: in each run $n_{i}$ samples are drawn at random from the training set for various values of $n_{i}$, $i = 1,\ldots,C$, and the signal covariances are estimated. The error rate of the MMAP classifier is then evaluated on the testing set. At the same time, we also determine if sufficient conditions for perfect classification as in Theorem 2 hold.  We run 1000 experimental runs with the Hopkins dataset, where in each run dataset is split at random into training and testing sets. We run 20 experimental runs with the MNIST dataset, where in each run the draw of the $n_i$ samples from the training set is random for $i = 1, \ldots, C$. 

The particular choice of samples in the training set can lead to high variability in the mismatched models, especially for small number of training samples. Therefore, in the following, we have chosen to report the results as follows:
\begin{itemize}
	\item  we state that analysis predicts reliable classification if the sufficient conditions in Theorem 2 hold with probability $p_p$ over the different experiment runs;
	\item we also state that simulation predicts reliable classification if the true error probability is 0 with probability $p_p$ over different experiment runs;
	\item if the simulated error rate exhibits an error floor we report the worst case error rate with probability $p_p$: the error rate that is achieved at least with probability $p_p$ over all experimental runs. 
\end{itemize}

\subsection{Results} \label{sec:application_results}

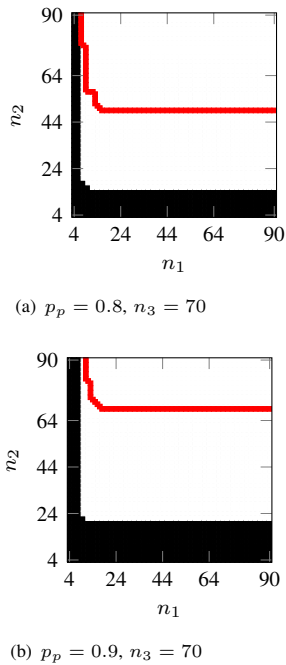
\begin{figure}[t]	
	\centering
	\subfigure[$p_{p} = 0.8$, $n_{3} = 70$]{
		\renewcommand{\LW}{1.5}
		\pgfplotsset{
		    matrix plot/.style={axis on top,clip marker paths=true,scale only axis,
        		height=\matrixrows/\matrixcols*\pgfkeysvalueof{/pgfplots/width},
       		enlarge x limits={rel=0.5/\matrixcols},
            enlarge y limits={rel=0.5/\matrixrows},
            scatter/use mapped color={draw=mapped color, fill=mapped color},
            scatter, point meta=explicit,mark=square*,
        		cycle list={mark size=0.5*\pgfkeysvalueof{/pgfplots/width}/\matrixcols}
    			},
		    matrix rows/.store in=\matrixrows,
		    matrix rows=44,
		    matrix cols/.store in=\matrixcols,
		    matrix cols=44,
		    colormap={blackwhite}{[2pt] 
		    rgb(0000pt)=(1,1,1);
		    rgb(1000pt)=(1,1,1);
		    rgb(1000pt)=(1,0,0);
		    rgb(2000pt)=(1,0,0);
		    rgb(2000pt)=(0,0,0);
		    rgb(3000pt)=(0,0,0);
			},
			}
		\begin{tikzpicture}[font=\footnotesize]    
			\begin{axis}[color=black,width=0.15\textwidth, matrix plot, align =center,
				    xtick = {1, 11,  21,  31, 44}, xticklabels = {4, 24, 44,  64, 90},
				    ytick = {1, 11,  21,  31, 44}, yticklabels = {4, 24, 44,  64, 90},
				    xlabel ={$n_{1}$}, ylabel ={$n_{2}$},	    
	        ] 
		    \addplot table [meta=val] {figdata/err08.txt};
		     \addplot table [meta = val] {figdata/errub08.txt};
		    \end{axis}    
	   \end{tikzpicture}
	} 
	\subfigure[$p_{p} = 0.9$, $n_{3} = 70$]{
		\renewcommand{\LW}{1.5}
		\pgfplotsset{
		    matrix plot/.style={axis on top,clip marker paths=true,scale only axis,
        		height=\matrixrows/\matrixcols*\pgfkeysvalueof{/pgfplots/width},
       		enlarge x limits={rel=0.5/\matrixcols},
            enlarge y limits={rel=0.5/\matrixrows},
            scatter/use mapped color={draw=mapped color, fill=mapped color},
            scatter, point meta=explicit,mark=square*,
        		cycle list={mark size=0.5*\pgfkeysvalueof{/pgfplots/width}/\matrixcols}
    			},
		    matrix rows/.store in=\matrixrows,
		    matrix rows=44,
		    matrix cols/.store in=\matrixcols,
		    matrix cols=44,
		    colormap={blackwhite}{[2pt] 
		    rgb(0000pt)=(1,1,1);
		    rgb(1000pt)=(1,1,1);
		    rgb(1000pt)=(1,0,0);
		    rgb(2000pt)=(1,0,0);
		    rgb(2000pt)=(0,0,0);
		    rgb(3000pt)=(0,0,0);
			},
			}
		\begin{tikzpicture}[font=\footnotesize]    
			\begin{axis}[color=black,width=0.15\textwidth, matrix plot, align =center,
				    xtick = {1, 11,  21,  31, 44}, xticklabels = {4, 24, 44,  64, 90},
				    ytick = {1, 11,  21,  31, 44}, yticklabels = {4, 24, 44,  64, 90},
				    xlabel ={$n_{1}$}, ylabel ={$n_{2}$},	    
	        ] 
		    \addplot table [meta=val] {figdata/err09.txt};
		     \addplot table [meta = val] {figdata/errub09.txt};
		    \end{axis}    
	   \end{tikzpicture}
	}
		\vspace*{-7pt}
	\caption{Phase transition of true error rate and phase transition given by the upper bound to the error probability as a function of number of training samples $n_1$, $n_2$. Black corresponds to an error floor of the true error rate, white corresponds to reliable classification, and red line denotes the phase transition predicted via Theorem 2 for a given probability $p_{p}$.} \label{fig:hopkins}
		\vspace*{-7pt}
\end{figure}

The results for the Hopkins dataset are reported in Fig. \ref{fig:hopkins}.

We observe that the phase transition predicted by analysis approximates reasonably well the phase transition obeyed by simulation. In particular, we can use our theory to gauge the number of training samples required for perfect classification in the low-noise regime. As expected, we also observe that the larger value of $p_{p}$ gives more conservative estimates of the required training samples. This holds for both simulation and analysis. 

We also observe that identical trends hold for other values of $n_3$. In particular, for $n_3 < 30$ simulation does not show a phase transition and likewise analysis does not show a phase transition either (these experiments are not reported in view of space limitations). In contrast, for $n_3 \geq 30$ both simulation and analysis predict a phase transition in the error probability.

\begin{figure*}[t]	
	\centering
	\subfigure[Classification of 2 digits]{
		\renewcommand{\LW}{1}
		\begin{tikzpicture}[font=\footnotesize]
		    \begin{axis}[color=black,height=0.25\textwidth, ylabel={Worst case error rate}, xlabel= {Training samples per class $n_\star$}, xmin = 25, xmax = 5000, 
		    ymin = 0, ymax = 0.02,
		    scaled y ticks = true, scaled y ticks=base 10:2, 
		    xtick={1000,3000,5000} ,   
		    xticklabels={$10^3$,$3 \cdot 10^3$,$5 \cdot 10^3$}   	
		]  
	    \addplot[color = red, mark=,line width=\LW] table [x index=0, y index=1] {figdata/mnist_error_2class.txt};
       \addplot[color = red, mark=,line width=2, dashed] coordinates {(1875, 0) (1875, 0.025)};
        \addplot[color = blue, mark=,line width=\LW] table [x index=0, y index=2] {figdata/mnist_error_2class.txt};
       \addplot[color = blue, mark=,line width=2, dashed] coordinates {(2625, 0) (2625, 0.025)};        
	    \end{axis} 
	\end{tikzpicture}
	}
	\subfigure[Classification of 5 digits]{
		\renewcommand{\LW}{1}
		\begin{tikzpicture}[font=\footnotesize]
		    \begin{axis}[color=black,height=0.25\textwidth, ylabel={Worst case error rate}, xlabel= {Training samples per class $n_\star$}, xmin = 25, xmax = 5000, 
		    ymin = 0, ymax = 0.1,
			scaled y ticks = true, scaled y ticks=base 10:2, 
		    xtick={1000,3000,5000} ,   
		    xticklabels={$10^3$,$3 \cdot 10^3$,$5 \cdot 10^3$}   	
		]  
	    \addplot[color = red, mark=,line width=\LW] table [x index=0, y index=1] {figdata/mnist_error_5class.txt};
	\addplot[color = red, mark=,line width=2, dashed] coordinates {(3225, 0) (3225, 0.11)};
        \addplot[color = blue, mark=,line width=\LW] table [x index=0, y index=2] {figdata/mnist_error_5class.txt};
       \addplot[color = blue, mark=,line width=2, dashed] coordinates {(3575, 0) (3575, 0.11)};        
	    \end{axis} 
	\end{tikzpicture}
	}	
	\subfigure[Classification of 10 digits]{
		\renewcommand{\LW}{1}
		\begin{tikzpicture}[font=\footnotesize]
		    \begin{axis}[color=black,height=0.25\textwidth, ylabel={Worst case error rate}, xlabel= {Training samples per class $n_\star$}, xmin = 25, xmax = 5000, 
		    ymin = 0, ymax = 0.1,
		    scaled y ticks = true, scaled y ticks=base 10:2, 
		    xtick={1000,3000,5000} ,   
		    xticklabels={$10^3$,$3 \cdot 10^3$,$5 \cdot 10^3$}   		    
		]  
	    \addplot[color = red, mark=,line width=\LW] table [x index=0, y index=1] {figdata/mnist_error_10class.txt};
       \addplot[color = red, mark=,line width=2, dashed] coordinates {(3375, 0) (3375, 0.11)};
        \addplot[color = blue, mark=,line width=\LW] table [x index=0, y index=2] {figdata/mnist_error_10class.txt};
       \addplot[color = blue, mark=,line width=2, dashed] coordinates {(3575, 0) (3575, 0.11)};   
	    \end{axis} 
	\end{tikzpicture}
	}		
		\vspace*{-7pt}
	\caption{The worst case error rate and phase transition predicted via Theorem 2 for a given probability $p_p$ are plotted for classification of MNIST digits. Solid red and blue lines correspond to worst case error rates for $p_p = 0.9$ and $p_p = 1$, respectively. Dashed vertical lines denote the phase transition predicted via Theorem 2 for $p_{p} = 0.9$ (red) and $p_p= 1$ (blue). } \label{fig:mnist}
		\vspace*{-7pt}
\end{figure*}
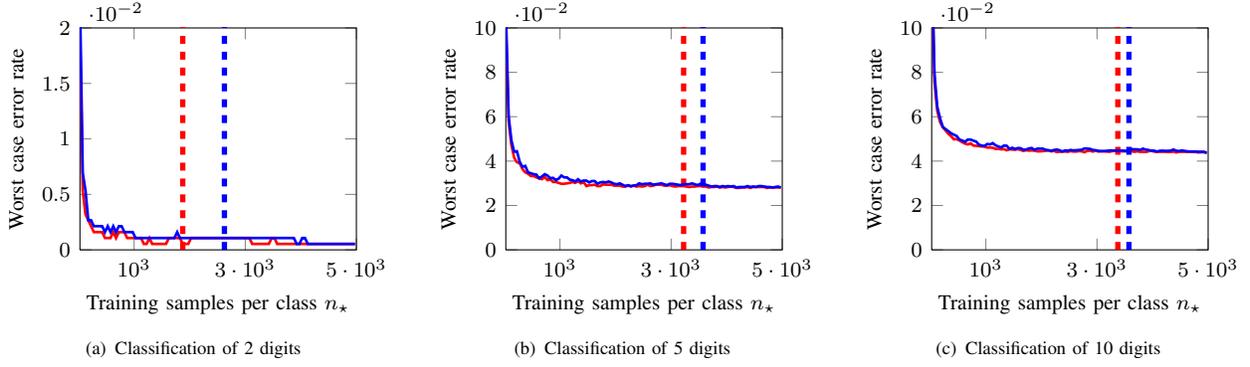

\review{
The results for the MNIST dataset are reported in Fig. \ref{fig:mnist}. Note that the number of training samples per class is the same for all classes, i.e. $n_i = n_\star$, $i = 1,\ldots,C$.}

\review{
In contrast to the results with the Hopkins dataset, the error rate obtained on the MNIST dataset exhibits an error floor. However, we observe that the worst case error rate reduces with a higher number of training samples and reaches an error floor at sufficiently large number of training samples. We also observe that the phase transition obtained via Theorem 2 predicts reasonably well the number of training samples needed to reach the error floor. }
		
Finally, note that real data are not drawn from Gaussian distributions or perfect linear subspaces (the two main ingredients underlying our analysis). Nevertheless, we have shown that the derived bound has practical value even when the two assumptions do not hold strictly.

\section{Conclusion}  \label{sec:conclusions}

This paper studies the classification of linear subspaces with mismatched classifiers, i.e. classifiers that operate on a mismatched version of the signal parameters \emph{in lieu} of the true signal parameters. In particular, we have developed a low-noise expansion of an upper bound to the error probability of such a mismatched classifier that equips one with a set of sufficient conditions -- which are a function of the geometry of the true signal distributions, the geometry of the mismatched signal distributions, and their interplay -- in order to understand whether it is possible to classify reliably in the presence of mismatch in the low-noise regime.

Such sufficient conditions are shown to be sharp in the sense that they can predict the presence (and the absence) of a classification error floor both in experiments involving synthetic data as well as experiments involving real data. These conditions have also been shown to gauge well the number of training samples required for reliable classification in a motion segmentation application using the Hopkins 155 dataset and a hand-written digit classification application using the MNIST dataset.

Overall, we argue that our conditions can also be used as a proxy to develop linear feature extraction methods that are robust to mismatch. In particular, our study suggests that such methods ought to orthogonalize the different classes as much as possible in order to tolerate model mismatch.  This intuition has been pursued in recent state-of-the-art linear feature extraction methods.


\section*{Appendix}

\subsection{Preliminaries} \label{sec:app_preliminaries}

We introduce additional quantities and Lemmas that are useful for the proofs.

\paragraph{Quantities}

 We define the projection operators:
\begin{IEEEeqnarray}{c}
	\bP_{i} = \bU_{i} \bU_{i}^{T} \,, \quad	\tbP_{i} = \tbU_{i} \tbU_{i}^{T}   \\
	\tbP_{ij}' = \tbU_{ij}' (\tbU_{ij}')^{T}	 \,, \label{eq:piprime}
\end{IEEEeqnarray}	
where $\bU_{i}$, $\tbU_{i}$ and $\tbU_{ij}'$ are given as in Section \ref{sec:geom}. In addition to the bases $\bU_{i}$ and $\tbU_{i}$ for the $\im(\bSigma_{i})$ and $\im(\tbSigma_{i})$, respectively, we also introduce the bases for the $\ke(\bSigma_{i})$ and $\ke(\tbSigma_{i})$ as $\bU_{i}^{\perp} \in \Rt{N}{N-r_{i}}$  and $\tbU_{i}^{\perp} \in \Rt{N}{N-\ti{r}_{i}}$, respectively. We define the projection operators onto this subspaces:
\begin{IEEEeqnarray}{rcl}
	\bK_{i} = \bU_{i}^{\perp} (\bU_{i}^{\perp})^{T} \,, \quad 
	 \tbK_{i} = \tbU_{i}^{\perp�} (\tbU_{i}^{\perp})^{T}\,. \label{eq:projK}
\end{IEEEeqnarray}
We also define 
 \begin{IEEEeqnarray}{rcl}
		\bL_{i} &=& \bU_{i} (\diag(\lambda_{1}^{i},\ldots, \lambda_{r_{i}}^{i}) +  \sn \bI)^{-1} (\bU_{i})^{T}  \\
		\tbL_{i} &=& \tbU_{i} (\diag(\ti{\lambda}_{1}^{i},\ldots, \ti{\lambda}_{\ti{r}_{i}}^{i}) + \sn \bI)^{-1} (\tbU_{i})^{T}	 �\,,
\end{IEEEeqnarray} 
and write
\begin{IEEEeqnarray}{rCl}
	\bSigma_{ij} &=& \bL_{ij} + \oneo{\sn} \bK_{ij} \,,
\end{IEEEeqnarray}
where $ 	\bL_{ij}  =  \bL_i + \alpha_{ij} \tbL_{j} - \alpha_{ij} \tbL_i$ and \mbox{$\bK_{ij}  =  \bK_i + \alpha_{ij} \tbK_{j} - \alpha_{ij} \tbK_i$}. 
Note that 
\begin{IEEEeqnarray}{rCl}
	 \bK_{ij} 
	 &=& \bK_i + \alpha_{ij} \tbP_{i} - \alpha_{ij} \tbP_{j} \\
	 &=& \bK_i + \alpha_{ij} \tbP_{ij}' - \alpha_{ij} \tbP_{ji}' \,	 
\end{IEEEeqnarray}
in view of the fact that $\bP_{i} + \bK_{i} = \bI$ and $\tbP_{i} + \tbK_{i} = \bI$ and $\tbP_{i} - \tbP_{j} = \tbP_{ij}' - \tbP_{ji}'$. The last equality simply follows from the definition of $ \tbP_{ij}' $ and $ \tbP_{ji}'$, and the definitions of $\tbU_{ij}'$, $\tbU_{ji}'$  and $\tbU_{ij}^{\cap}$ given in Section  \ref{sec:geom}:
\begin{IEEEeqnarray}{rCl}
	\tbP_{i} - \tbP_{j} &=& {\tbU_{ij}}'({\tbU_{ij}}')^{T} + \tbU_{ij}^{\cap}(\tbU_{ij}^{\cap})^{T} \nonumber \\
	& & - \left( {\tbU_{ji}}'({\tbU_{ji}}')^{T} + \tbU_{ij}^{\cap}(\tbU_{ij}^{\cap})^{T} \right) \\
	&=& \tbP_{ij}'  - \tbP_{ji}'\,.
	\label{eq:diffP}
\end{IEEEeqnarray}

Finally, we present a decomposition of $\bx \in \Rn$. We write 
\begin{equation}
\bx = \bx_{\parallel} + \bx_{\perp} = \bx_{\bV} + \bx_{\bW} + \bx_{\perp},
\end{equation}
where 
\begin{IEEEeqnarray}{rCl}
		 \bx_{\parallel} &= & \bU_{i} \bz_{\parallel} 	\\
		 \bx_{\perp} & = & \bU_{i}^{\perp} \bz_{\perp}\label{eq:xdecomp3} 	\\
		 \bx_{\bV} & =  &\bV_{ij} \bz_{\bV}  \label{eq:xdecomp1} \\
		 \bx_{\bW} &= & \bW_{ij} \bz_{\bW}  \label{eq:xdecomp2}, 
\end{IEEEeqnarray}
for some vectors $\bz_{\parallel}  \in \mathbb{R}^{r_i} ,\bz_{\perp} \in \mathbb{R}^{N-r_i},\bz_{\bV} \in \mathbb{R}^{s^V_{ij}}$ and $\bz_{\bW}\in \mathbb{R}^{s^W_{ij}} $. Note also that $\| \bx_{\bV} \| = \| \bz_{\bV} \|$, $\| \bx_{\bW} \| = \| \bz_{\bW} \|$ and $\| \bx_{\perp} \| = \| \bz_{\perp} \|$. 

\paragraph{Lemmas}
\begin{lemma} \label{th:kernel_tbPij_prime}
The following equality holds:
\begin{IEEEeqnarray}{c}
	\im(\tbU_{ij}')^\perp = \ke(\tbP_{ij}') = \ke(\tbU_{i}) + (\im(\tbU_{i}) \cap \im(\tbU_{j}))  \,.\nonumber 
\end{IEEEeqnarray}
\begin{IEEEproof}
By leveraging the definition of $\tbP_{ij}'$ in \eqref{eq:piprime} we have
\begin{IEEEeqnarray}{rCl}
\ke(\tbP_{ij}')  & = & \left( \im(\tbP_{ij}')   \right)^\perp = \left( \im({\tbU_{ij}}')   \right)^\perp = \im([\tbU_{ij}^{\cap}, {\tbU_{i}^{\perp}}]) \IEEEeqnarraynumspace \nonumber \\
 & = & \im(\tbU_{i})^\perp + (\im(\tbU_{i}) \cap \im(\tbU_{j})). \nonumber
\end{IEEEeqnarray}
\end{IEEEproof}
\end{lemma}

\begin{lemma} \label{th:pa_posdef_lemma}
The following statement holds:
\begin{IEEEeqnarray}{rCl}
	d_{\min}(\bV_{ij}, \tbU_{ij}') &<& d_{\max}(\bV_{ij}, \tbU_{ji}') \label{eq:pa_posdef_1} \\
	&\implies & \nonumber \\	
	\bV_{ij}^{T} (\tbU_{ij}' (\tbU_{ij}')^{T} &-& \tbU_{ji}' (\tbU_{ji}')^{T}) \bV_{ij} \succ  \bZero \,. \label{eq:pa_posdef_2}
\end{IEEEeqnarray}
\begin{IEEEproof}
First, note that  
\begin{IEEEeqnarray*}{rCl}
\bV_{ij}^{T} (\tbU_{ij}' (\tbU_{ij}')^{T} - \tbU_{ji}' (\tbU_{ji}')^{T}) \bV_{ij} =(\bV_{ij})^{T} (\tbP_{ij}' - \tbP_{ji}') \bV_{ij}\,.
\end{IEEEeqnarray*}
Then we write the following
\begin{IEEEeqnarray}{l}
	(\bV_{ij})^{T}  \tbP_{ij}' \bV_{ij} =(\bV_{ij})^{T} \tbU_{ij}' (\tbU_{ij}')^{T} \bV_{ij} \\
	(\bV_{ij})^{T}  \tbP_{ji}' \bV_{ij} =(\bV_{ij})^{T} \tbU_{ji}' (\tbU_{ji}')^{T} \bV_{ij} .
\end{IEEEeqnarray}
Note that the singular values of $(\bV_{ij})^{T} \tbU_{ij}'$ and $(\bV_{ij})^{T} \tbU_{ji}'$ correspond to the cosines of the principal angles between and $\im(\bV_{ij})$ and $\im(\tbU_{ij}')$, and $\im(\bV_{ij})$ and $\im(\tbU_{ji}')$, respectively. We then consider the SVDs 
\begin{IEEEeqnarray}{c}
	(\bV_{ij})^{T} \tbU_{ij}' = \bH_{ij} \bD_{ij} \bJ_{ij}^{T} \\
	(\bV_{ij})^{T} \tbU_{ji}' = \bH_{ji} \bD_{ji} \bJ_{ji}^{T}
\end{IEEEeqnarray}
where the dimensions of matrices $\bH_{ij}$, $\bH_{ji}$, $\bD_{ij}$, $\bD_{ji}$, $\bJ_{ij}$ and $\bJ_{ji}$ follow from the dimension of the $\bV_{ij}$, $\tbU_{ij}'$ and $\tbU_{ji}'$ as shown in \eqref{eq:svdPA}. We can now express \eqref{eq:pa_posdef_2} as
\begin{IEEEeqnarray}{c}
	\bH_{ij} \bD_{ij} \bD_{ij}^{T} \bH_{ij}^{T} \succ \bH_{ji} \bD_{ji} \bD_{ji}^{T} \bH_{ji}^{T}.
	 \label{eq:suffSVD}
\end{IEEEeqnarray}
It is straightforward to see that \eqref{eq:pa_posdef_1} implies \eqref{eq:pa_posdef_2}.
\end{IEEEproof}
\end{lemma}
\begin{lemma}
The following equalities and inequalities hold:
\begin{IEEEeqnarray}{rCl}
	\bx^{T} \bL_{i} \bx  & \geq & \oneo{\lambda_{1}^{i} + 1} \| \bx_{\parallel} \|^{2} \label{eq:ineq1} \\
 	 \bx^{T} ( \tbL_{j} - \tbL_{i}) \bx  & \geq & -  \oneo{\ti{\lambda}_{\ti{r}_{i}}^{i} } \| \bx \|^{2} \label{eq:ineq3} \\
  	\bx^{T} \bK_{i} \bx   & = &  \| \bx_{\perp} \|^{2} \,. \label{eq:ineq4}
\IEEEeqnarraynumspace
\end{IEEEeqnarray}
\begin{IEEEproof}
The inequality in \eqref{eq:ineq1} is due to the fact that $\bx_{\parallel} \in \im(\bL_{i}) = \im(\bU_{i})$ and $\oneo{\lambda_{1}^{i} + 1}$ is a lower bound to the minimum positive eigenvalue of $\bL_{i}$. 
The inequality in \eqref{eq:ineq3} is due to the fact that $\tbL_{j}$ is positive semidefinite and that $\oneo{\ti{\lambda}_{\ti{r}_{i}}^{i} }$ is an upper bound for the largest eigenvalue of $\tbL_{i}$.  The equality in \eqref{eq:ineq4} follows from the definition of the projector $\bK_i$. 
\end{IEEEproof}
\end{lemma}
\begin{lemma}
Assume that 
	\begin{IEEEeqnarray}{c}
	\im(\bW_{ij}) \subseteq \im(\tbU_{ji}')^\perp \text{ and } \label{eq:neccproof_lemma} \\
	\bV_{ij}^{T} (\tbU_{ij}' (\tbU_{ij}')^{T} - \tbU_{ji}' (\tbU_{ji}')^{T}) \bV_{ij} \succ \bZero \,. \label{eq:suffproof_lemma}
\end{IEEEeqnarray}
Denote by $c_{0}$ the smallest eigenvalue of 
\begin{IEEEeqnarray*}{rCl}
	\bV_{ij}^{T} (\tbU_{ij}' (\tbU_{ij}')^{T} - \tbU_{ji}' (\tbU_{ji}')^{T}) \bV_{ij} \\ =(\bV_{ij})^{T} (\tbP'_{i} - \tbP'_{j}) \bV_{ij} \,.
\end{IEEEeqnarray*}
Then 
\begin{IEEEeqnarray}{rCl}
\bx^{T} (\tbK_{j}  - \tbK_{i}) \bx    \geq   c_{0}  \| \bx_{\bV} \|^{2}   - 2 \| \bx_{\bV} \|  \| \bx_{\perp} \|  -  \| \bx_{\perp} \|^{2} \,. \nonumber \\ \label{eq:ineq5}
\end{IEEEeqnarray}
\begin{IEEEproof}
Note that \eqref{eq:xdecomp2} implies $\bx_{\bW} \in \im(\bW_{ij}) = \im(\bSigma_{i}) \cap \ker(\tbP_{ij}')$, and the condition \eqref{eq:suffproof_lemma} also implies $\bx_{\bW} \in \ker(\tbP_{ji}') = \im(\tbU_{ji}')^\perp$. Then, we can write
\begin{IEEEeqnarray}{rCl}
	\bx^{T} (\tbK_{j}  - \tbK_{i}) \bx & = & \bx^{T} (\tbP_{ij}'  - \tbP_{ji}') \bx \\
	 &=& \bx_{\bV}^{T} (\tbP_{ij}' - \tbP_{ji}') \bx_{\bV} \nonumber \\
	 & & + 2 \bx_{\bV}^{T} (\tbP_{ij}' - \tbP_{ji}') \bx_{\perp} \IEEEeqnarraynumspace  \nonumber \\
	& &+  \bx_{\perp}^{T} (\tbP_{ij}' - \tbP_{ji}') \bx_{\perp}  
\end{IEEEeqnarray}
and we note that condition \eqref{eq:suffproof_lemma} implies the lower bound $ \bx_{\bV}^{T} (\tbP_{ij}' - \tbP_{ji}') \bx_{\bV} \geq c_{0} \| \bx_{\bV} \|^{2}$. Moreover, all the eigenvalues of  $\tbP_{ij}' - \tbP_{ji}'$ are contained in \mbox{ the interval $[-1,1]$ \cite[Theorem 26]{Galantai2008}}, so that $ \bx_{\perp}^{T} (\tbP_{ij}' - \tbP_{ji}') \bx_{\perp}  \geq -  \| \bx_{\perp} \|^{2}$, and, on leveraging Cauchy-Schwarz inequality, we also have $\bx_{\bV}^{T} (\tbP_{ij}' - \tbP_{ji}') \bx_{\perp} \geq - 2 \| \bx_{\bV} \|  \| \bx_{\perp} \|$.
\end{IEEEproof}
\end{lemma}

\subsection{Proof of Theorem 1}
We prove Theorem 1 by using the fact that $u(x) \leq \exp(\alpha x)$, $\forall x, \alpha > 0$ and by leveraging the union bound. 

Recall from \eqref{eq:MMAPerror} that the error probability associated with the MMAP classifier can be expressed as 
\begin{IEEEeqnarray}{c}
	P(e) = \sum_{i=1}^{C} p_{i} \cdot P(e | c = i)
\end{IEEEeqnarray}
where $P(e | c = i) = P(\hat{c} \neq i | c =  i)$ is the error probability for signals in class $i$. Via the union bound, we can state that
\begin{IEEEeqnarray}{c}
	P(e | c = i) = P(\hat{c} \neq i | c =  i) \leq \sum_{j=1, j\neq i}^{C} P(\hat{c} = j | c =  i) \IEEEeqnarraynumspace
\end{IEEEeqnarray}
where
\begin{IEEEeqnarray}{rcl}
	P(\hat{c} = j | c =  i)  &=& \Iii  p(\by|c=i) \\
					& & \cdot u\left(\log \left( \frac {\ti{p}_{j} \ti{p}(\by | c = j)} {\ti{p}_{i} \ti{p}(\by | c = i) } \right) \right) \, \dd\by \,.
\end{IEEEeqnarray}
We will denote $P(\hat{c} = j | c =  i)  = P(e_{ij})$.  Now, by letting $\alpha_{ij} > 0 \, \forall i \neq j$ we can upper bound the step function to obtain
\begin{IEEEeqnarray}{rCl}
	P(e_{ij}) 
	& \leq & \Iii  p(\by|c=1) \nonumber \\
	& & \cdot \exp\left(\alpha_{ij} \log \left(\frac{\ti{p}_{j} \ti{p}(\by | c = j)}{\ti{p}_{i} \ti{p}(\by | c = i)}\right)\right) \, \dd\by \nonumber \\
	& = &  \left(\frac{\ti{p}_j }{\ti{p}_i}\right)^{\alpha_{ij}} \left(\frac{|\tbSigma_{i} + \sn \bI|}{|\tbSigma_{j} + \sn \bI|}\right)^{\frac{\alpha_{ij}}{2}}  \nonumber \\
	& & \cdot ((2 \pi)^{N} |\bSigma_i + \sn \bI|)^{-\oneo{2}} \nonumber \\
	& & \cdot \Iii \exp\left(-\frac{1}{2} \by^{T} \bSigma_{ij} \by \right)	 \dd\by  = \bar{P}(e_{ij}) , \nonumber \\*
	\label{eq:pe12proof}
\end{IEEEeqnarray}
where we recall
\begin{equation*}
\bSigma_{ij} =  (\bSigma_{i} + \sn \bI)^{-1}  + \alpha_{ij} (\tbSigma_{j} + \sn \bI)^{-1}  
		  - \alpha_{ij} (\tbSigma_{i} + \sn \bI)^{-1} \,.
\end{equation*}

If $\bSigma_{ij} \succ \bZero\, \forall i \neq j $, then the integral in \eqref{eq:pe12proof} converges \mbox{$\forall i \neq j$}. Therefore, we can bound the error probability as follows:
	\begin{IEEEeqnarray}{c}
		P(e) \leq \bar{P}(e) = \sum_{i=1}^{C} p_{i} \cdot \left( \sum_{j = 1, j \neq i}^{C} \bar{P}(e_{ij}) \right) 
	\end{IEEEeqnarray}
	where
		\begin{IEEEeqnarray}{rCl}
			\bar{P}(e_{ij}) &=&  \left(\frac{\ti{p}_j }{\ti{p}_i} \sqrt{\frac{|\tbSigma_{i} + \sn \bI|}{|\tbSigma_{j} + \sn \bI|}} \right)^{\alpha_{ij}} \cdot (|\bSigma_i + \sn \bI| |\pmmSigma{i}{j}|)^{-\oneo{2}} \,.  \nonumber \\ 
		\end{IEEEeqnarray}

If $\exists i \neq j : \bSigma_{ij} \not\succ \bZero$ then the integral in \eqref{eq:pe12proof} does not converge. Therefore, we trivially bound the error probability as $	P(e) \leq  \bar{P}(e) \leq 1 \,.$
		 
\subsection{Proof of Theorem 2}


The proof is presented in two parts. First, we establish sufficient conditions for $\bSigma_{ij} \succ \bZero$; second, we establish conditions for the upper bound to the probability of misclassification to approach zero as the noise approaches zero.

\subsubsection{Positive Definiteness of $\bSigma_{ij}$} \label{sec:proofpart1}

The following two Lemmas gives sufficient conditions for $\bSigma_{ij} \succ \mathbf{0}$. 
\begin{lemma} \label{th:sufficient_conditions1}
Assume that $s^V_{ij} > 0$,
\begin{IEEEeqnarray}{c}
	\im(\bW_{ij}) \subseteq \im(\tbU_{ji}')^\perp \,,  \label{eq:neccproof} \\
	\bV_{ij}^{T} (\tbU_{ij}' (\tbU_{ij}')^{T} - \tbU_{ji}' (\tbU_{ji}')^{T}) \bV_{ij} \succ \bZero \,, \label{eq:suffproof} \\
		\alpha_{ij} <  \min \left(\frac{\ti{\lambda}^{i}_{\ti{r}_{i}}}{\lambda_{1}^{i} + 1}, \frac{c_{0}}{1 + c_{0} (1 + \oneo{\ti{\lambda}^{i}_{\ti{r}_{i}}})}, 1 \right) \,,\label{eq:suffaproof}
\end{IEEEeqnarray}
where $c_{0}$ is the smallest eigenvalue of 
\begin{IEEEeqnarray*}{rCl}
	\bV_{ij}^{T} (\tbU_{ij}' (\tbU_{ij}')^{T} - \tbU_{ji}' (\tbU_{ji}')^{T}) \bV_{ij} \\ =(\bV_{ij})^{T} (\tbP'_{i} - \tbP'_{j}) \bV_{ij} \,.
\end{IEEEeqnarray*}

Then 
\begin{IEEEeqnarray}{rCl}
	\bSigma_{ij} \succ \mathbf{0}, \, \forall \sigma^2 \in \left(0, \min \left(1, \frac{1 - \alpha_{ij}}{\alpha_{ij}} \ti{\lambda}_{i}^{\ti{r}_{i}} \right) \right)\,. \label{eq:lemma_posdef_1}
\end{IEEEeqnarray}
\begin{IEEEproof}
	To show this we first produce a lower bound:
	\begin{IEEEeqnarray}{rCl}
\bx^T \bSigma_{ij} \bx & = & \bx^T \bL_{ij} \bx + \frac{1}{\sigma^2}  \bx^T \bK_{ij} \bx  \label{eq:sigijdecomp}\\
\nonumber
 & = & \bx^T \bL_i \bx + \alpha_{ij}   \bx^T ( \tbL_{j} -  \tbL_i) \bx \\
 & & +  \frac{1}{\sigma^2}  \left(     \bx^T \bK_i \bx + \alpha_{ij}   \bx^T ( \tbK_{j} -  \tbK_i) \bx     \right) \\
 & \geq &  \bc^{T} \bC \bc,
\end{IEEEeqnarray}
where 
$\bc = [\| \bx_{\bW} \|, \| \bx_{\bV} \|, \| \bx_{\perp} \|]^{T}$ 
and
\begin{equation}
 \bC =  \left[ \begin{array}{ccc}  \oneo{\lambda_{1}^{i} + 1} -  \frac{\alpha_{ij}}{\ti{\lambda}^{i}_{\ti{r}_{i}}} & 0 & 0 \\ 0 & \oneo{\lambda_{1}^{i} + 1} -  \frac{\alpha_{ij}}{\ti{\lambda}^{i}_{\ti{r}_{i}}} + \frac{c_{0}\alpha_{ij}}{\sn}  & - \frac{\alpha_{ij}}{\sn} \\ 0 & - \frac{\alpha_{ij}}{\sn} &  \frac{1 - \alpha_{ij}}{\sn} -   \frac{\alpha_{ij}}{\ti{\lambda}^{i}_{\ti{r}_{i}}} \end{array} \right],
\end{equation}
by using the equalities and inequalities \eqref{eq:ineq1}-\eqref{eq:ineq4} and \eqref{eq:ineq5}.

 Now, by using standard algebraic manipulations, it is possible to show that the choice \eqref{eq:suffaproof} leads to $\bC \succ \mathbf{0}$	, hence \eqref{eq:lemma_posdef_1} holds.
	
\end{IEEEproof}
\end{lemma}
\begin{lemma} \label{th:sufficient_conditions2}
	Assume that $s^V_{ij}=0$,
	\begin{IEEEeqnarray}{c}
	\im(\bW_{ij}) \subseteq \im(\tbU_{ji}')^\perp \,, \\
		\alpha_{ij} <   \min \left(\frac{\ti{\lambda}^{i}_{\ti{r}_{i}}}{\lambda_{1}^{i} + 1}, \frac{\ti{\lambda}^{i}_{\ti{r}_{i}}}{\ti{\lambda}^{i}_{\ti{r}_{i}} + 1},1 \right) \,.  
		\label{eq:suffa2proof}
\end{IEEEeqnarray}
	Then
	\begin{IEEEeqnarray}{rCl}
		\bSigma_{ij} \succ \mathbf{0}, \, \forall \sigma^2 \in \left(0, \min \left(1, \frac{1 - \alpha_{ij}}{\alpha_{ij}} \ti{\lambda}_{i}^{\ti{r}_{i}} \right) \right)\,. \label{eq:lemma_posdef_2} 
	\end{IEEEeqnarray}
\begin{IEEEproof}
	We prove the Lemma by constructing the lower bound 
	\begin{IEEEeqnarray}{rCl}
	\bx^{T} \bSigma_{ij} \bx  &\geq& \oneo{\lambda_{1}^{i} + 1} \| \bx_{\parallel} \|^{2} -  \frac{\alpha_{ij}}{\ti{\lambda}_{\ti{r}_{i}}^{i} } \| \bx \|^{2}  \nonumber \\  & &+ \oneo{\sn} \left( \| \bx_{\perp} \|^{2}  + \alpha_{ij} \bx^{T} (\tbP_{ij}'  - \tbP_{ji}') \bx  \right)  \\
	&\geq&  \left(\oneo{\lambda_{1}^{i} + 1} - \frac{\alpha_{ij}}{\ti{\lambda}^{i}_{\ti{r}_{i}} } \right)\|\bx_{\parallel} \|^{2}  \nonumber \\
	& & +\left(\frac{1-\alpha_{ij}}{\sn} - \frac{\alpha_{ij}}{\ti{\lambda}^{i}_{\ti{r}_{i}} } \right) \|\bx_{\perp} \|^{2}\,,
\end{IEEEeqnarray}
by using the inequalities equalities and inequalities \eqref{eq:ineq1}-\eqref{eq:ineq4} and \eqref{eq:ineq5}, and by noting that $\bx_{\bV} = \bZero$. The choice \eqref{eq:suffa2proof} then leads to \eqref{eq:lemma_posdef_2}.
\end{IEEEproof}
	
\end{lemma}

\subsubsection{Part 2: Low-noise Expansion}

To obtain the low-noise expansion of the upper bound to the error probability we first present two supporting Lemmas.
\begin{lemma} \label{th:Kij_semiposdef}
Assume that condition \eqref{eq:neccproof} given in Lemma \ref{th:sufficient_conditions1} holds. Assume also that $s_{ij}^V > 0$ and  \eqref{eq:suffproof} and \eqref{eq:suffaproof} given in Lemma \ref{th:sufficient_conditions1} hold,  or that $s_{ij}^V = 0$ and \eqref{eq:suffa2proof} given in Lemma \ref{th:sufficient_conditions2} holds. Then $\bK_{ij} \succeq \bZero$ and $\rank(\bK_{ij}) = N + s^V_{ij} - r_{i}$.
\begin{IEEEproof}
Assume that \eqref{eq:neccproof}, $s^V_{ij}>0$ \eqref{eq:suffproof} and \eqref{eq:suffaproof} are satisfied. By definition, $\im(\bW_{ij}) =  \im(\bSigma_{i}) \cap \ke(\tbP_{ij}')$ and, as a consequence of (\ref{eq:neccproof}), it also holds $\im(\bW_{ij}) \subseteq \ke(\tbP_{ji}')$, which leads to $\im(\bW_{ij}) \subseteq \ke(\bK_{ij})$. Moreover, it is straightforward to note that $\im([\bV_{ij}, \bU_{i}^{\perp}]) = (\im(\bW_{ij}))^{\perp}$. Then, in order to prove that $\bK_{ij} \succeq \bZero$, we show that $\bx^T \bK_{ij} \bx=\bx^{T} (\bK_{i} + \alpha_{ij} ( \tbP_{ij}' - \tbP_{ji}') ) \bx > 0, \forall \bx \in \im([\bV_{ij}, \bU_{i}^{\perp}])$. Namely, by leveraging the equality in \eqref{eq:ineq4} and inequality in \eqref{eq:ineq5}, we can write
\begin{IEEEeqnarray}{rCl}
	\bx^{T} (\bK_{i} + \alpha_{ij} ( \tbP_{ij}' - \tbP_{ji}') ) \bx &\geq& (1-\alpha_{ij}) \| \bx_{\perp} \|^{2} \nonumber \\
	& &- 2 \alpha_{ij} \| \bx_{\perp} \| \| \bx_{\bV} \| \nonumber \\ 
	& &+\alpha_{ij} c_{0} \| \bx_{\bV} \|^{2}  \,, \label{eq:proofineq1}
\end{IEEEeqnarray}
where $\bx_{\perp}, \bx_{\bV}$ have been defined in \eqref{eq:xdecomp3} and \eqref{eq:xdecomp1}. If $\alpha_{ij} < \frac{c_{0}}{c_{0} + 1}$ then the right hand side of \eqref{eq:proofineq1} is always strictly positive, unless $\mathbf{x}=\mathbf{0}$. Then, since the condition in \eqref{eq:suffaproof} implies $\alpha_{ij} < \frac{c_{0}}{c_{0} + 1}$, we can conclude that $\bK_{ij} \succeq \bZero$ and $\im(\bW_{ij}) = \ke(\bK_{ij})$ and $\im([\bV_{ij}, \bU_{i}^{\perp}]) = \im(\bK_{ij})$. Therefore, $\rank(\bK_{ij}) = \rank([\bV_{ij}, \bU_{i}^{\perp}]) = s^V_{ij} + (N - r_{i})$.

Assume now that \eqref{eq:neccproof}, $s^V_{ij}=0$ and \eqref{eq:suffa2proof} are satisfied. In this case $ \bx^{T} \bK_{ij} \bx  =\bx^{T} (\bK_{i} + \alpha_{ij} ( \tbP_{ij}' - \tbP_{ji}') ) \bx = \bx_{\perp}^{T} (\bK_{i} + \alpha_{ij} ( \tbP_{ij}' - \tbP_{ji}') ) \bx_{\perp} \geq \| \bx_{\perp} \|^{2} (1 -\alpha_{ij})$, where we have used the fact that eigenvalues of $\tbP_{ji}' - \tbP_{ji}'$ contained in the interval $[-1,1]$. Since $\eqref{eq:suffa2proof}$ implies $\alpha_{ij} < 1$ we conclude, via an argument similar to that in previous paragraph, that $\bK_{ij} \succeq \bZero$ and $\rank(\bK_{ij}) = \rank(\bU_{i}^{\perp}) = s^V_{ij} + (N - r_{i})$. 

\end{IEEEproof}
\end{lemma}
\begin{lemma} \label{th:Sigma_ij_expansion}
Assume that condition \eqref{eq:neccproof} given in Lemma \ref{th:sufficient_conditions1} holds. Assume also that $s_{ij}^V > 0$ and  \eqref{eq:suffproof} and \eqref{eq:suffaproof} given in Lemma \ref{th:sufficient_conditions1} hold,  or that $s_{ij}^V = 0$ and \eqref{eq:suffa2proof} given in Lemma \ref{th:sufficient_conditions2} holds. Then, as $\sn \to 0$, we can write
\begin{equation}
| \bL_{ij} + \oneo{\sn} \bK_{ij}| = v_{ij} \cdot \left(  \frac{1}{\sn}  \right)^{r_{\bK_{ij}}}  + \mathcal{O} \left( \left(  \frac{1}{\sn}  \right)^{r_{\bK_{ij}}-1} \right),
\label{eq:detLK}
\end{equation}
where $r_{\bK_{ij}}=\rank(\bK_{ij})$, and  $v_{ij}$ is given as
\begin{IEEEeqnarray}{c}
	v_{ij} =  \left\{ \begin{array}{ll} \pdet(\bK_{ij}) | ( \bU_{\bK_{ij}}^{\perp})^{T} \bL^{0}_{ij}  \bU_{\bK_{ij}}^{\perp} | &\text{if } r_{\bK_{ij}} < N \\ 
							|\bK_{ij}| &\text{if } r_{\bK_{ij}} = N   \end{array}  \right. \,, \nonumber \\* \label{eq:v12proof}
\end{IEEEeqnarray}
$\bL^{0}_{ij} = \lim_{\sn \rightarrow 0} \bL_{ij} = \bL_{i}^{0} + \alpha_{ij} \tbL_{j}^{0}  - \alpha_{ij} \tbL_{i}^{0}$ and
 \begin{IEEEeqnarray}{rcl}
		&&\bL^{0}_{i} = \bU_{i} (\diag(\lambda_{1}^{i},\ldots, \lambda_{r_{i}}^{i}))^{-1} (\bU_{i})^{T}  \\
		&& \tbL^{0}_{i} = \tbU_{i} (\diag(\ti{\lambda}_{1}^{i},\ldots, \ti{\lambda}_{\ti{r}_{i}}^{i}))^{-1} (\tbU_{i})^{T} \,.
\end{IEEEeqnarray} 

\begin{IEEEproof}
Note first that the sufficient conditions imply $\bK_{ij} \succeq \bZero$ via Lemma \ref{th:Kij_semiposdef}. We can write the eigenvalue decomposition of $\bK_{ij}$:
\begin{IEEEeqnarray}{c}
	\bK_{ij} = \bU_{\bK_{ij}} \left[ \begin{array}{cc} \bLambda_{\bK_{ij}} & \bZero \\ \bZero & \bZero \end{array} \right] \bU_{\bK_{ij}}^{T}\,,
\end{IEEEeqnarray}
where $\bU_{\bK_{ij}} \in \Rt{N}{N}$ is orthogonal and $\bLambda_{\bK_{ij}} = \diag(\lambda_{1}^{\bK_{ij}}, \ldots, \lambda^{\bK_{ij}}_{r_{\bK_{ij}}})$ contains the positive eigenvalues of $\bK_{ij}$, with $r_{\bK_{ij}}=\rank(\bK_{ij})$.

Now, we can write
\begin{IEEEeqnarray}{rCl}
	|\bL_{ij} + \oneo{\sn} \bK_{ij}|  &=&  \left| \left[ \begin{array}{cc} \oneo{\sn} \bLambda_{\bK_{ij}} & \bZero \\ \bZero & \bZero \end{array} \right]+ \bE \right | \,,
\end{IEEEeqnarray}
where $\bE= \bU_{\bK_{ij}}^T \bL_{ij}  \bU_{\bK_{ij}} $. 
We also denote by $\bE_{i_{1} \ldots i_{m}}$ the principal submatrix of order $N-m$ obtained by deleting the rows and the columns $i_{1}, \ldots, i_{m}$
 of the matrix $\bE$. Note that $\bE_{i_{1} \ldots i_{m}} = \bP_{i_{1} \ldots i_{m}}^{T} \bE \bP_{i_{1} \ldots i_{m}}$, where the matrix $\bP_{i_{1} \ldots i_{m}} \in \Rt{N}{N-m}$ is obtained by picking all the columns from the identity matrix with the column indices different from $i_{1}, \ldots, i_{m}$. Then, the Poincar\'e separation theorem \cite[Corollary 4.3.37]{Horn2012} guarantees that the eigenvalues $\bE_{i_{1} \ldots i_{m}}$ are bounded by the minimum and the maximum eigenvalues of $\bE$, which correspond to the minimum and maximum eigenvalues of $\bL_{ij}$. Moreover, as $\sn \to 0$, while the diagonal elements of $ \oneo{\sn} \bLambda_{\bK_{ij}}$ grow unbounded, the eigenvalues of $\bL_{ij}$, and therefore, also the determinant of $\bE_{i_{1} \ldots i_{m}}$, are bounded. 
 
 Then, we can use the determinant decomposition in \mbox{\cite[Theorem 2.3]{Ipsen2008}} to express $|\bL_{ij} + \oneo{\sn} \bK_{ij}|$ as follows. If $r_{\bK_{ij}} = N$:
 \begin{IEEEeqnarray}{c}
	|\bL_{ij} + \oneo{\sn} \bK_{ij}| = |\oneo{\sn} \bK_{ij}| +  |\bL_{ij}| + S_{1} + \ldots + S_{N-1} \,, \label{eq:decomp1} \nonumber \\*
\end{IEEEeqnarray}
where
\begin{IEEEeqnarray}{rCl}
	S_{m} $=$ \sum_{1 \leq i_{1} < \ldots < i_{m} \leq N} \left(\oneo{\sn} \lambda_{i_{1}}^{\bK_{ij}} \right) \cdots \left(\oneo{\sn} \lambda_{i_{m}}^{\bK_{ij}} \right) |\bE_{i_{1} \ldots i_{m}}|\, \nonumber \\
	$$	1 \leq m \leq N-1 \, \nonumber \\* \label{eq:Sm1}
\end{IEEEeqnarray}
and the summation is over all possible ordered subsets of $m$ indices from the set $\{1, \ldots, r_{\bK_{ij}} \}$.
Otherwise, if $r_{\bK_{ij}} < N$:
 \begin{IEEEeqnarray}{c}
	|\bL_{ij} + \oneo{\sn} \bK_{ij}| = |\bL_{ij}| + S_{1} + \ldots  + S_{r_{\bK_{ij}}} \,, \label{eq:decomp2}
\end{IEEEeqnarray}
where 
\begin{IEEEeqnarray}{rCl}
	S_{m} $=$ \sum_{1 \leq i_{1} < \ldots < i_{m} \leq r_{\bK_{ij}}} \left(\oneo{\sn} \lambda_{i_{1}}^{\bK_{ij}} \right) \cdots \left(\oneo{\sn} \lambda_{i_{m}}^{\bK_{ij}} \right) |\bE_{i_{1} \ldots i_{m}}|\, \nonumber \\
	$$	1 \leq m \leq r_{\bK_{ij}} \,. \nonumber \\* \label{eq:Sm2}
\end{IEEEeqnarray}
Now we show that \eqref{eq:v12proof} holds. We first assume $\rank(\bK_{ij}) = N$ and take the right hand side of \eqref{eq:decomp1} and multiply it by $\left(  \frac{1}{\sn}  \right)^{r_{\bK_{ij}}} \left( {\sn}  \right)^{r_{\bK_{ij}}}$ to get 
\begin{IEEEeqnarray}{l}
	\left(  \frac{1}{\sn}  \right)^{r_{\bK_{ij}}} \left( |\bK_{ij} | + \left(  {\sn}  \right)^{r_{\bK_{ij}}} \left( |\bL_{ij}| + S_{1} + \ldots + S_{N-1} \right) \right) \,. \nonumber \\ \*
\end{IEEEeqnarray}
Note now that for all $S_{m}$, $m = 1, \ldots, N-1$, $\lim_{\sn \to 0} (\sn)^{r_{\bK_{ij}}} S_{m} = 0$ and $\lim_{\sn \to 0} (\sn)^{r_{\bK_{ij}}} |\bL_{ij}| = 0$. Therefore,  \eqref{eq:v12proof} holds for the case $\rank(\bK_{ij}) = N$. To show the derivation of $v_{ij}$ for the case $\rank(\bK_{ij}) < N$ we use the same technique where we multiply by $\left(  \frac{1}{\sn}  \right)^{r_{\bK_{ij}}} \left( {\sn}  \right)^{r_{\bK_{ij}}}$ the right hand side of \eqref{eq:decomp2} to get
\begin{IEEEeqnarray}{rCl}
	\left(  \frac{1}{\sn}  \right)^{r_{\bK_{ij}}} \biggl(   &&   \left(  {\sn}  \right)^{r_{\bK_{ij}}} S_{r_{\bK_{ij}}} + \nonumber \\ && \left(  {\sn}  \right)^{r_{\bK_{ij}}} \left( |\bL_{ij}| + S_{1} + \ldots + S_{r_{\bK_{ij}} -1} \right) \biggl) \,. \nonumber \\*
\end{IEEEeqnarray}
As $\sn \to 0$ we can write $ \left(  {\sn}  \right)^{r_{\bK_{ij}}} S_{r_{\bK_{ij}}}  = \pdet(\bK_{ij} ) | ( \bU_{\bK_{ij}}^{\perp})^{T} \bL^{0}_{ij}  \bU_{\bK_{ij}}^{\perp} | $. This concludes the derivation of \eqref{eq:v12proof}. Note also that $v_{ij}>0$, since the pseudo-determinant and the determinants in \eqref{eq:v12proof} are greater than zero. 

\end{IEEEproof}
\end{lemma}

We now provide the low-noise expansion of  the upper bound to the probability of misclassification.

Assume that sufficient conditions for positive definiteness of $\bSigma_{ij}, \forall i \neq j$ do not hold. Then, the upper bound to the probability of error is chosen to be $\bar{P} (e) = 1$, so that in general it does not tend to zero as $\sigma^2$ tends to zero.

Assume now that the sufficient conditions for $\bSigma_{ij} \succ \bZero$ as given in the first part of the proof hold $\forall i \neq j$. Then, the upper bound to the probability of misclassification can be written as follows:\footnote{	Note that a value for which  $\alpha_{ij}$ satisfies the conditions for $\bSigma_{ij} \succ \bZero$ always exists and therefore does not affect the derivation of the low-noise expansion.}
\begin{IEEEeqnarray}{rCl}
					\bar{P}(e) = \sum_i \sum_{j \neq i} p_i & & \left(\frac{\ti{p}_j }{\ti{p}_i} \sqrt{\frac{|\tbSigma_{i} + \sn \bI|}{|\tbSigma_{j} + \sn \bI|}} \right)^{\alpha_{ij}} \nonumber \\ & & \cdot (|\bSigma_i + \sn \bI| |\pmmSigma{i}{j}|)^{-\oneo{2}} \,. 
					\label{eq:UBmulticlass}
		\end{IEEEeqnarray}	
We will now produce a low-noise expansion of \eqref{eq:UBmulticlass} in order to understand whether or not $\lim_{\sigma^2 \to 0} \bar{P} (e) = 0$. The following low-noise expansions are trivial:
\begin{IEEEeqnarray}{rCl}
			|{\bSigma}_{i} + \sn \bI |  &= & \left( \prod_{k=1}^{{r}_{i}} ({\lambda}^{i}_{k}  + \sn)\right)  \left(\sn \right)^{N- {r}_{i}}\label{eq:det1} \\
			&   = & \mathcal{O}   \left(  \left(\sn \right)^{N- {r}_{i}} \right),   \qquad \sn \to 0 \nonumber \\
			|\ti{\bSigma}_{i} + \sn \bI |&  = & \left( \prod_{k=1}^{\ti{r}_{i}} (\ti{\lambda}^{i}_{k} + \sn)\right)  \left(\sn \right)^{N- \ti{r}_{i}}  \label{eq:det1t} \\
			&   = & \mathcal{O}   \left(  \left(\sn \right)^{N- {\ti{r}}_{i}} \right) ,  \qquad \sn \to 0  \nonumber \,.
		\end{IEEEeqnarray}	
The low-noise expansion of $\displaystyle |\bSigma_{ij}|$
is more involved and it is provided in Lemma \ref{th:Sigma_ij_expansion}.

Then, it follows immediately that the low-noise expansion of each term in the upper bound to the probability of error in \eqref{eq:UBmulticlass} is given by
\begin{equation}
 A_{ij} \left( \sn \right)^{d_{ij}} + o\left(   \left(\sn \right)^{d_{ij}} \right), \label{eq:expansionPei}
\end{equation}
where 
\begin{IEEEeqnarray}{rCl}
	d_{ij} &=& -\frac{\alpha_{ij}}{2} \left((N- \ti{r}_{i}) - (N - \ti{r}_{j}) \right) \nonumber \\
	& &- \oneo{2} (N - r_{i}) - \oneo{2} (- \rank(\bK_{ij}))  \nonumber \\
	&= &  \oneo{2} (\alpha_{ij}(\ti{r}_{j} - \ti{r}_{i}) + s^V_{ij})	 \,,	\label{eq:dorderproof1}
\end{IEEEeqnarray}
\begin{IEEEeqnarray}{c}
	A_{ij} = \left(\frac{\ti{p}_{j}}{\ti{p}_{i}} \right)^{\alpha_{ij}}  \left( \frac{\ti{v}_{i}}{\ti{v}_{j}} \right)^{\frac{\alpha_{ij}}{2}} \left( v_{i} v_{ij}   \right)^{-\oneo{2}} >0
\end{IEEEeqnarray}
and 
\begin{IEEEeqnarray}{l}
	v_{i} = \pdet(\bSigma_{i})  \,, \quad \ti{v}_{i} = \pdet(\tbSigma_{i})  \,.	
\end{IEEEeqnarray}
It follows immediately that the low-noise expansion of the upper bound to the probability of error in \eqref{eq:UBmulticlass} is given by
\begin{IEEEeqnarray}{c}
	\bar{P}(e) = A  \left({\sn} \right)^{d} + o\left(   \left(\sn \right)^{d} \right)\,, \label{eq:expansionPe}
\end{IEEEeqnarray}
where $d = \min_{(i \neq j)}d_{ij}$ and $A=\sum_{(i,j) \in \mathcal{S}_d} A_{ij}$ where  $\mathcal{S}_d = \{  (i,j) : d_{ij} =d \}$.

\subsection{Proof of Corollary 2}
Assume $s^V_{ij} > 0 \, \forall (i,j), i \neq j$ and 
\begin{IEEEeqnarray}{c}
	d_{\min}(\bU_{i}, \tbU_{i}) <  d_{\max}(\bU_{i}, \tbU_{j}) \, \forall (i,j),  i \neq j \,.
\end{IEEEeqnarray}
Note that $d_{\min}(\bU_{i}, \tbU_{i}) <  d_{\max}(\bU_{i}, \tbU_{j})$ implies
\begin{IEEEeqnarray}{c}
	\bU_{i}^{T} (\tbU_{i} \tbU_{i}^{T} - \tbU_{j} \tbU_{j}^{T} )\bU_{i} \succ \bZero \\
	 \iff \nonumber \\
		\bU_{i}^{T} (\tbU_{ij}' (\tbU_{ij}')^{T} - \tbU_{ji}' (\tbU_{ji}')^{T} )\bU_{i}  \succ \bZero  \,,\label{eq:col2last}
\end{IEEEeqnarray}
where we have used result in Lemma \ref{th:pa_posdef_lemma} in the Appendix A.

By taking $\bx \in \bW_{ij}$ or $\bx \in \bV_{ij}$ it is straightforward to show that \eqref{eq:col2last} implies \eqref{eq:nec} and \eqref{eq:suff}, thus obtaining conditions identical to those in Corollary 1.

\subsection{Proof of Corollary 3}
We prove the corollary by showing that in diagonal case \eqref{eq:suff} always holds. Note first that
\begin{IEEEeqnarray*}{rCl}
\im(\bV_{ij}) &=& \im(\bU_i) \cap ( \im (\bU_i) \cap \im(\tbU_{ij}')^\perp )^\perp \\
 &=& \im(\bU_i) \cap \im(\tbU_{ij}') \subseteq \im(\tbU_{ij}') \,.
\end{IEEEeqnarray*}
It is also straightforward to establish that \eqref{eq:suff} holds if and only if $\im(\bV_{ij}) \subseteq \im(\tbU_{ji}')^\perp$, and this always holds since $\im(\bV_{ij}) \subseteq \im(\tbU_{ij}'$ and $\im(\tbU_{ij}' \subseteq \im(\tbU_{ji}')^\perp$.

\subsection{Derivation of Example \ref{th:example_perturbation}}
We prove statement \eqref{eq:suffpert}, by showing that 
\begin{IEEEeqnarray}{c}
	1 - \delta_{12} >  N (\epsilon_{1} + \epsilon_{2})
\end{IEEEeqnarray}
together with $s_{12}, s_{21} > 0$ implies the sufficient conditions for perfect classification in Corollary 2.

Assume $\bU_{i}$ and $\bU_{j}$ are given and the singular values of $(\bU_{i})^{T} \bU_{j}$ are known. We also know that $\tbU_{j} = \bQ_{j} \bU_{j}$. We can write
\begin{IEEEeqnarray}{c}
	(\bU_{i})^{T} \tbU_{j} = (\bU_{i})^{T} \bU_{j} + (\bU_{i})^{T} (\bQ_{j} - \bI) \bU_{j}
\end{IEEEeqnarray}
On leveraging \cite[Theorem 1]{Stewart1998}, we can state that the $i$-th singular value $\ti{d}_{i}$ associated with $(\bU_{i})^{T} \tbU_{j}$ lies in the interval $[d_{i} - \| (\bU_{i})^{T} (\bQ_{j} - \bI) \bU_{j}\|_{2}, d_{i} + \|(\bU_{i})^{T} (\bQ_{j} - \bI) \bU_{j} \|_{2}]$, where $d_{i}$ is the $i$-th singular value of $(\bU_{i})^{T} \bU_{j}$. Then, we can write the upper bound
\begin{IEEEeqnarray}{c}
	\|(\bU_{i})^{T} (\bQ_{j} - \bI) \bU_{j} \|_{2} \leq \|  \bQ_{j} - \bI  \|_{2} = \epsilon_{j} \label{eq:SVbound}
	\end{IEEEeqnarray}
where the first inequality follows from the SVD separation theorem \cite[Theorem 2.2]{Rao1979}. Note also that
\begin{IEEEeqnarray}{c}
	(\bU_{i})^{T} \tbU_{i} = \bI + (\bU_{i})^{T} (\bQ_{i} - \bI) \bU_{i}
\end{IEEEeqnarray}
where the singular values of $(\bU_{i})^{T} \tbU_{i}$ are bounded by $1 \pm \|(\bU_{i})^{T} (\bQ_{i} - \bI) \bU_{i} \|_{2}$. By leveraging \eqref{eq:SVbound} we can further bound the singular values as $1 \pm  \epsilon_{i}$.

Note now that $1 -\delta_{12} > (\epsilon_{1} + \epsilon_{2})$ if and only if $1 -  \epsilon_{1} >  \delta_{12}  +  \epsilon_{2}$, which implies
\begin{IEEEeqnarray}{rCl}
	d_{\min}(\bU_{1}, \tbU_{1}) &<&  d_{\max}(\bU_{1}, \tbU_{2})\,, \label{eq:distiff1}
\end{IEEEeqnarray}
and is also equivalent to
\begin{IEEEeqnarray}{rCl}
	\max_{k} \cos_{k}( (\bU_{1})^{T} \tbU_{2}) &<& \min_{l} \cos_{l}( (\bU_{1})^{T} \tbU_{1}) \label{eq:distiff2}
	\,,
\end{IEEEeqnarray}
where $\max_{k} \cos_{k}( (\bU_{1})^{T} \tbU_{2}) $ denotes the cosine of the smallest principal angle between $\im(\bU_{1})$ and $\im(\tbU_{2})$,  $\max_{k} \cos_{k}( (\bU_{1})^{T} \tbU_{2}) $ denotes the cosine of the largest principal angle between $\im(\bU_{1})$ and $\im(\tbU_{1})$. The equivalence between \eqref{eq:distiff1} and \eqref{eq:distiff2} follows straight from the definition of min and max correlation distances. It is now easy to verify that $1 -  \epsilon_{1} >  \delta_{12}  +  \epsilon_{2}$ implies \eqref{eq:distiff2}, since $1 -  \epsilon_{1}$ is a lower bound for the cosine of the largest principal angles between $\bU_{1}$ and $\tbU_{1} $, and  $ \delta_{12}  +  \epsilon_{2}$ is an upper bound to the cosine of the smallest principal angles between $\bU_{1}$ and $\tbU_{2}$. 

Finally, the same arguments can be used to show that $	d_{\min}(\bU_{2}, \tbU_{2}) <  d_{\max}(\bU_{2}, \tbU_{1})$. This concludes the proof.



\Urlmuskip=0mu plus 1mu
\bibliographystyle{style/jureIEEEtran}
\bibliography{IEEEabrv,style/libraryISIT2015}


\ifCLASSOPTIONcaptionsoff
 \newpage
\fi

%

\begin{IEEEbiography}[{\includegraphics[width=1in,height=1.25in,clip,keepaspectratio]{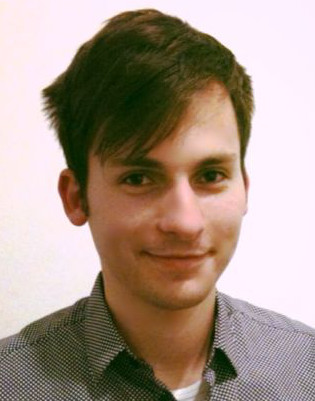}}]{Jure Sokoli{\' c}}
(S'14) received his Diploma in Electrical Engineering from University of Ljubljana in 2013. Currently, he is working towards his Ph.D. in the Department of Electrical \& Electronic Engineering at University College London. His research interest focus on high-dimensional data processing and machine learning. 
\end{IEEEbiography}
\begin{IEEEbiography}[{\includegraphics[width=1in,height=1.25in,clip,keepaspectratio]{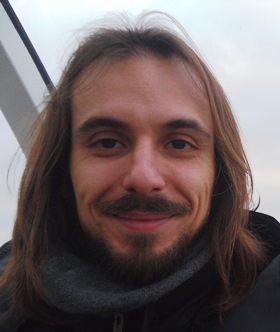}}]{Francesco Renna}
(S'09--M'11) received his Laurea Specialistica Degree in Telecommunication Engineering and Ph.D. degree in Information Engineering, both from University of Padova, in 2006 and 2011, respectively. Between 2007 and 2015 he held visiting researcher and postdoctoral appointments at Infineon Technology AG, Princeton University, Georgia Institute of Technology (Lorraine Campus), Sup{\'e}lec, University of Porto, Duke University and University College London. Since 2016 he is a Marie Curie fellow at University of Cambridge. His research interests focus on high-dimensional information processing but also include physical layer security for multicarrier and multiantenna systems.
\end{IEEEbiography}
\begin{IEEEbiography}[{\includegraphics[width=1in,height=1.25in,clip,keepaspectratio]{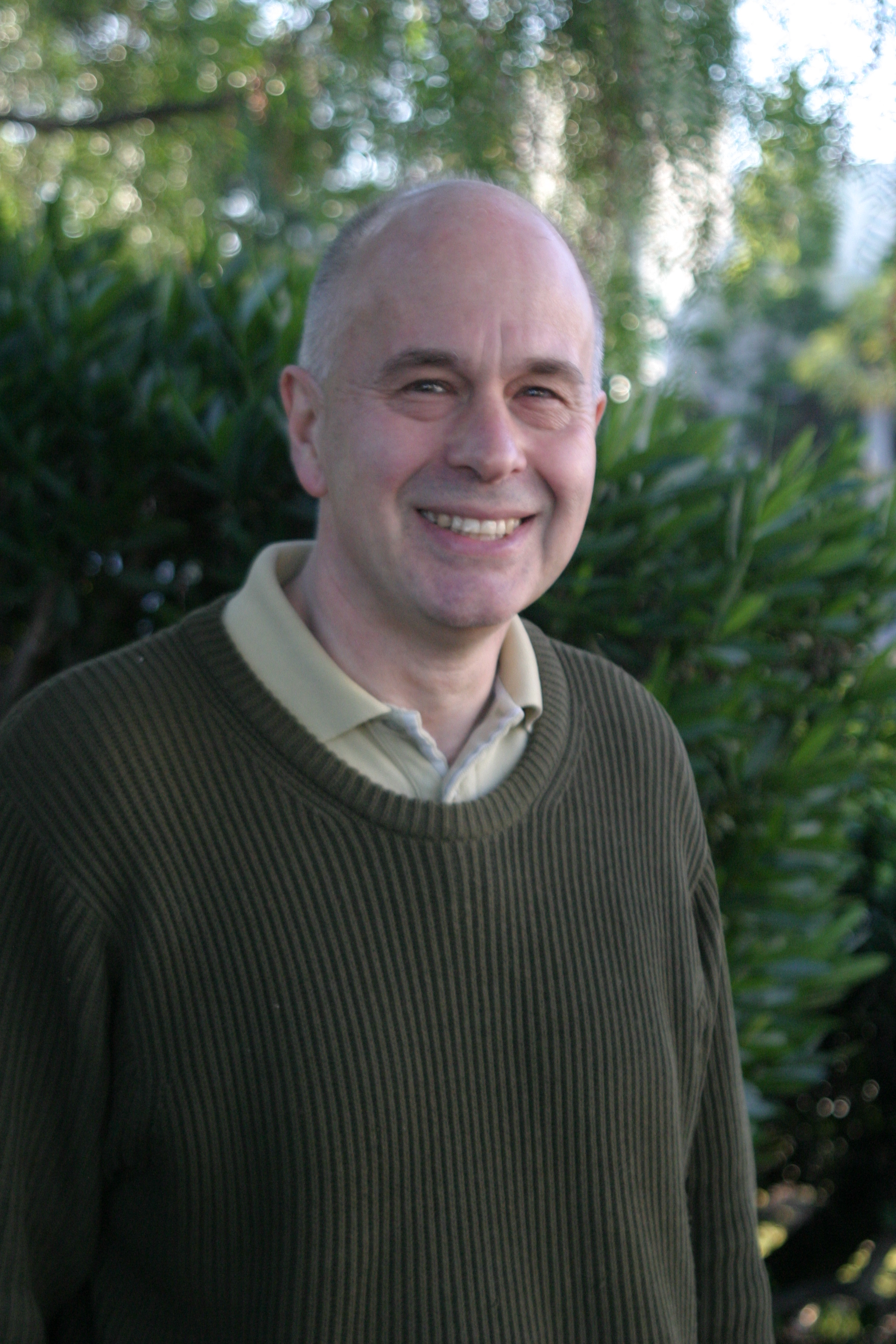}}]{Robert Calderbank}
(M'89--SM'97--F'98) received the BSc degree in 1975 from Warwick University, England, the MSc degree in 1976 from Oxford University, England, and the PhD degree in 1980 from the California Institute of Technology, all in mathematics.

Dr. Calderbank is Professor of Electrical Engineering at Duke University where he now directs the Information Initiative at Duke (iiD) after serving as Dean of Natural Sciences (2010-2013). Dr. Calderbank was previously Professor of Electrical Engineering and Mathematics at Princeton University where he directed the Program in Applied and Computational Mathematics. Prior to joining Princeton in 2004, he was Vice President for Research at AT\&T, responsible for directing the first industrial research lab in the world where the primary focus is data at scale.  At the start of his career at Bell Labs, innovations by Dr. Calderbank were incorporated in a progression of voiceband modem standards that moved communications practice close to the Shannon limit. Together with Peter Shor and colleagues at AT\&T Labs he showed that good quantum error correcting codes exist and developed the group theoretic framework for quantum error correction. He is a co-inventor of space-time codes for wireless communication, where correlation of signals across different transmit antennas is the key to reliable transmission.
 
Dr. Calderbank served as Editor in Chief of the IEEE TRANSACTIONS ON INFORMATION THEORY from 1995 to 1998, and as Associate Editor for Coding Techniques from 1986 to 1989. He was a member of the Board of Governors of the IEEE Information Theory Society from 1991 to 1996 and from 2006 to 2008. Dr. Calderbank was honored by the IEEE Information Theory Prize Paper Award in 1995 for his work on the Z4 linearity of Kerdock and Preparata Codes (joint with A.R. Hammons Jr., P.V. Kumar, N.J.A. Sloane, and P. Sole), and again in 1999 for the invention of space-time codes (joint with V. Tarokh and N. Seshadri). He has received the 2006 IEEE Donald G. Fink Prize Paper Award, the IEEE Millennium Medal, the 2013 IEEE Richard W. Hamming Medal, and he was elected to the US National Academy of Engineering in 2005.
\end{IEEEbiography}
\begin{IEEEbiography}[{\includegraphics[width=1in,height=1.25in,clip,keepaspectratio]{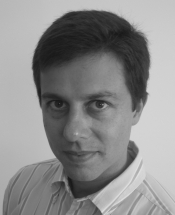}}]{Miguel R. D. Rodrigues}
 (S'98–-M'02) received the Licenciatura degree in Electrical Engineering from the Faculty of Engineering of the University of Porto, Portugal in 1998 and the Ph.D. degree in Electronic and Electrical Engineering from University College London, UK in 2002.
 
He is currently a Reader with the Department of Electronic and Electrical Engineering, University College London, UK. He was previously with the Department of Computer Science, University of Porto, Portugal, where he also led the Information Theory and Communications Research Group at Instituto de Telecomunica\c{c}\~{o}es Porto. He has also held visiting research appointments at Princeton University, USA, Duke University, USA, Cambridge University, UK and University College London, UK in the period 2007 to 2013. His research interests are in the general areas of information theory, communications theory and signal processing.

Dr. Rodrigues was the recipient of the IEEE Communications and Information Theory Societies Joint Paper Award in 2011 for the work on Wireless Information-Theoretic Security (with M. Bloch, J. Barros and S. W. McLaughlin). He was also the recipient of the the Prize Engenheiro Ant{\'o}nio de Almeida, the Prize Engenheiro Cristiano Spratley, and the Merit Scholarship from the University of Porto, and the best student poster prize at the 2nd IMA Conference on Mathematics in Communications. He was also awarded doctoral and postdoctoral research fellowships from the Portuguese Foundation for Science and Technology, and research fellowships from Foundation Calouste Gulbenkian.

Dr. Rodrigues is an associated editor to IEEE Communications Letters.
\end{IEEEbiography}

%
%
%




\end{document}